\documentclass[sigconf]{acmart}
\usepackage{subfigure}
\usepackage{multirow}
\usepackage{wrapfig}
\AtBeginDocument{%
  \providecommand\BibTeX{{%
    \normalfont B\kern-0.5em{\scshape i\kern-0.25em b}\kern-0.8em\TeX}}}

\copyrightyear{2022} 
\acmYear{2022} 
\setcopyright{acmcopyright}\acmConference[MM '22]{Proceedings of the 30th ACM International Conference on Multimedia}{October 10--14, 2022}{Lisboa, Portugal}
\acmBooktitle{Proceedings of the 30th ACM International Conference on Multimedia (MM '22), October 10--14, 2022, Lisboa, Portugal}
\acmPrice{15.00}
\acmDOI{10.1145/3503161.3547854}
\acmISBN{978-1-4503-9203-7/22/10}

\begin{document}

\title{SingGAN: Generative Adversarial Network For High-Fidelity Singing Voice Generation}


\author{Rongjie Huang$^{*}$}
\affiliation{%
  \institution{Zhejiang University}
  \country{}
  }
\email{rongjiehuang@zju.edu.cn}

\author{Chenye Cui$^{*}$}
\affiliation{%
  \institution{Zhejiang University}
  \country{}
  }
\email{chenyecui@zju.edu.cn}

\author{Feiyang Chen$^{*}$}
\affiliation{%
  \institution{Huawei Cloud}
  \country{}
  }
\email{chenfeiyang2@huawei.com}

\author{Yi Ren}
\affiliation{%
  \institution{Zhejiang University}
  \country{}
  }
\email{rayeren@zju.edu.cn}

\author{Jinglin Liu}
\affiliation{%
  \institution{Zhejiang University}
  \country{}
  }
\email{jinglinliu@zju.edu.cn}

\author{Zhou Zhao$^{\dagger}$}
\affiliation{%
  \institution{Zhejiang University}
  \country{}
  }
\email{zhaozhou@zju.edu.cn}

\author{Baoxing Huai}
\affiliation{%
  \institution{Huawei Cloud}
  \country{}
  }
\email{huaibaoxing@huawei.com}

\author{Zhefeng Wang}
\affiliation{%
  \institution{Huawei Cloud}
  \country{}
  }
\email{wangzhefeng@huawei.com}

\renewcommand{\shortauthors}{Rongjie Huang et al.}



\begin{abstract}
\renewcommand{\thefootnote}{}
\footnotetext{$^*$ Equal contribution. $\dagger$ Corresponding author.}
Deep generative models have achieved significant progress in speech synthesis to date, while high-fidelity singing voice synthesis is still an open problem for its long continuous pronunciation, rich high-frequency parts, and strong expressiveness. Existing neural vocoders designed for text-to-speech cannot directly be applied to singing voice synthesis because they result in glitches and poor high-frequency reconstruction. In this work, we propose SingGAN, a generative adversarial network designed for high-fidelity singing voice synthesis. Specifically, 1) to alleviate the glitch problem in the generated samples, we propose source excitation with the adaptive feature learning filters to expand the receptive field patterns and stabilize long continuous signal generation; and 2) SingGAN introduces global and local discriminators at different scales to enrich low-frequency details and promote high-frequency reconstruction; and 3) To improve the training efficiency, SingGAN includes auxiliary spectrogram losses and sub-band feature matching penalty loss. To the best of our knowledge, SingGAN is the first work designed toward high-fidelity singing voice vocoding. Our evaluation of SingGAN demonstrates the state-of-the-art results with higher-quality (MOS 4.05) samples. Also, SingGAN enables a sample speed of 50x faster than real-time on a single NVIDIA 2080Ti GPU. We further show that SingGAN generalizes well to the mel-spectrogram inversion of unseen singers, and the end-to-end singing voice synthesis system SingGAN-SVS enjoys a two-stage pipeline to transform the music scores into expressive singing voices. Audio samples are available at \url{https://SingGAN.github.io/}
\end{abstract}




\maketitle

\section{Introduction}

\begin{figure}[ht]
  \centering
   \vspace{-6mm}
  \includegraphics[width=0.5\textwidth]{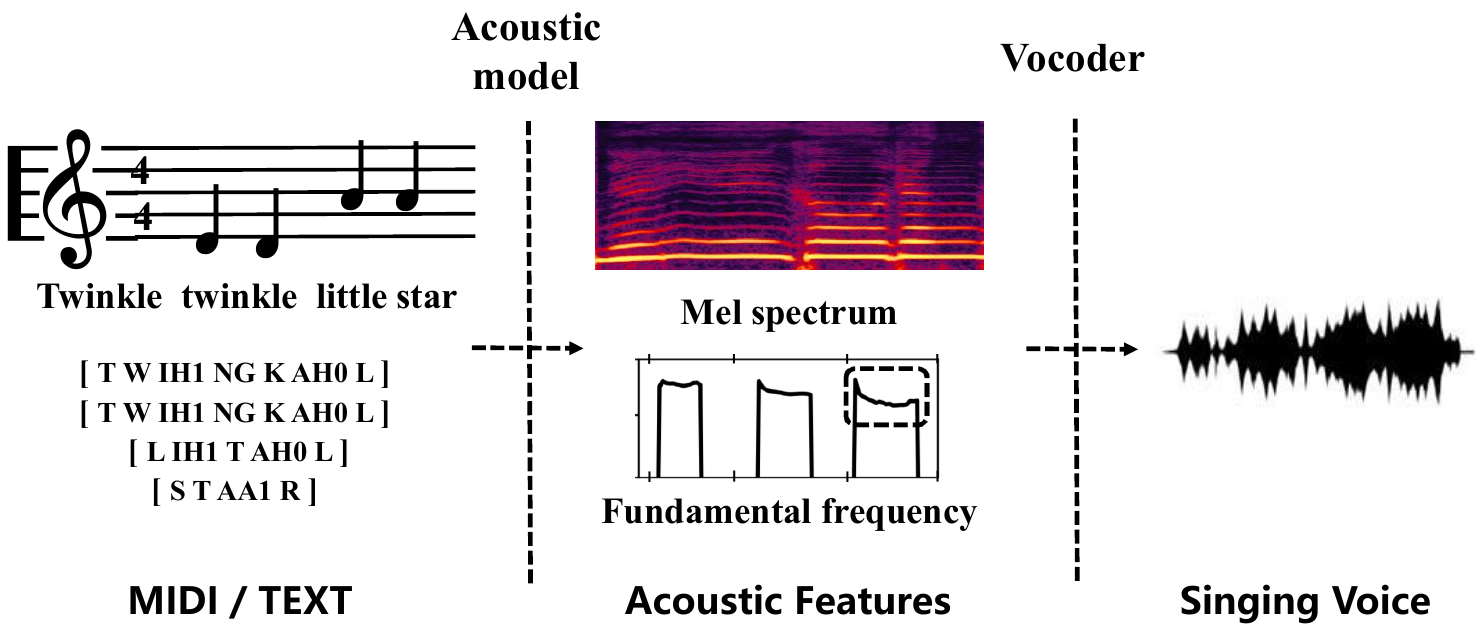}
    \vspace{-6mm}
 \caption{Singing voice synthesis overall pipeline.}
  \label{fig:svs}
   \vspace{-2mm}
\end{figure}

Singing voice synthesis (SVS) has been widely employed in various applications with human-computer interaction, such as virtual anchor, artificial intelligence singer, and recommendation system ~\cite{DBLP:conf/www/ZhangYYLFZC022,DBLP:conf/mm/ZhangTYZKLZYW20,DBLP:conf/kdd/ZhangTZYKJZYW20,huang2022transpeech}. As illustrated in Figure~\ref{fig:svs}, similar to text-to-speech (TTS) systems~\cite{Fastspeech,tacotron,huang2022fastdiff,cui2021emovie,huang2022generspeech}, SVS systems~\cite{gu2020bytesing,chen2020hifisinger,DeepSinger,liu2021diffsinger} generally adopt an acoustic model to convert the musical score~\footnote{A music score consists of lyrics, pitch and duration.} into the intermediate features (i.e., mel-spectrogram) and a vocoder to synthesize singing voices. In this work, we focus on designing a second-stage model that efficiently synthesizes high-fidelity waveforms from mel-spectrograms.

It is well-known that modeling singing voices could be more challenging than traditional speech, which typically holds 1) a longer continuous pronunciation; and 2) rich high-frequency parts for improving expressiveness. To the best of our knowledge, few vocoders have been explicitly designed for singing voice generation, and thus most previous SVS systems~\cite{gu2020bytesing,chen2020hifisinger} have to utilize speech vocoders to transform the generated mel-spectrograms into waveforms. Through preliminary analyses and revisits (see Section~\ref{analysis}), we find that distinct degradation of audio quality emerges when applying speech models in singing voice vocoding for the following reasons: 1) Speech vocoders usually lose attention and time dependencies in the long continuous pronunciation, and therefore glitches emerge in the generated samples; 2) Speech vocoders typically lack the capacity of reconstructing high-frequency signals, which leads to hissing noises and unnatural sounds in the high-frequency parts of generated samples.

In this paper, we propose SingGAN, a novel generative adversarial network for high-fidelity singing voice generation. Specifically, 1) SingGAN introduces the source excitation module with the adaptive feature learning filters to expand receptive fields and stabilize long continuous signal generation, efficiently reducing glitches in the generated singing voices. 2) SingGAN utilizes a global discriminator to rebuild overall characteristics (e.g., singer identity and prosody) and local discriminators at different scales to enrich low-frequency details and promote high-frequency reconstruction. 3) To improve the training efficiency of the generator and the fidelity of the generated audio, SingGAN introduces the sub-band feature matching penalty loss and auxiliary spectrogram losses. The joint training approach effectively works in GANs for singing voice vocoding. Based on SingGAN, we introduce the extension system SingGAN-SVS, which enjoys a two-stage pipeline to transform the music scores into expressive singing voices.

Experimental results show that SingGAN achieves the state-of-the-art results (MOS 4.05) and outperforms the best publicly available baselines. To the best of our knowledge, this is the first work designed toward high-fidelity singing voice vocoding. SingGAN, requiring only 1.59M parameters, can generate high-quality singing voices 50x faster than real-time on a single NVIDIA 2080Ti without engineered kernels. Our further experiments demonstrated its outperformed generalizability to unseen singers. SingGAN-SVS successfully simplifies the pipeline to generate high-fidelity singing voices from the music scores. 

The contributions of this paper are summarized as follows:
\begin{itemize}
  \item Considering the long continuous pronunciation, high sampling rate, and strong expressiveness of the singing voices, we analyze and revisit the glitches and poor high-frequency reconstruction when adopting previous speech models in singing voice vocoding.
  \item We propose SingGAN, a novel generative adversarial network with source excitation and adaptive feature learning filters. SingGAN utilizes the global and local discriminators in adversarial learning and adopts the auxiliary spectrogram loss with sub-band feature matching penalty objectives for waveform reconstruction. To the best of our knowledge, SingGAN is the first work designed for high-fidelity singing voice vocoding.
  \item Experimental results demonstrate that SingGAN achieves state-of-the-art results (MOS 4.05) in singing voice vocoding and outperforms the best publicly available neural vocoders. We further show that SingGAN generalizes well to the mel-spectrogram inversion of unseen singers, and the end-to-end singing voice synthesis system SingGAN-SVS could directly generate high-fidelity singing voices from lyrics.
\end{itemize}

Our audio samples are available on the demo website \footnote{\url{https://SingGAN.github.io/}}.
\section{Related Works}
In this section, we provide the research background of text-to-speech and singing voice synthesis and briefly review several variations of neural vocoders.

\subsection{Text-to-Speech}
Text-to-Speech (TTS)~\cite{tacotron,Fastspeech,popov2021gradtts} aims to synthesize natural and intelligible speech given text as input, which has witnessed significant progress in recent years. Previous neural TTS models~\cite{tacotron,li2019neural} first generate mel-spectrograms autoregressively from text and then synthesize speech from the generated mel-spectrograms using a separately trained vocoder~\cite{oord2016wavenet,kalchbrenner2018efficient}. Since autoregressive models sequentially generate a sample, they usually suffer from slow inference speed. Recently, several works~\cite{Fastspeech,kim2020glowtts} have been proposed to generate mel-spectrogram frames in parallel, which speed up mel-spectrogram generation and preserve perceptual quality. SVS systems are mostly inspired by TTS and follow the primary components, such as text-to-audio alignment, parametric acoustic modeling, and neural vocoding.

\subsection{Singing Voice Synthesis}
In recent years, neural network-based singing voice synthesis (SVS) has made rapid progress and attracted much attention in the machine learning and speech community. HiFiSinger~\cite{chen2020hifisinger} utilizes the acoustic model for the text-to-intermediate generation and a vocoder to synthesize singing voices from intermediate features. ByteSing~\cite{gu2020bytesing} is a Chinese SVS system based on duration prediction, which could generate natural singing voices given music scores. DiffSinger~\cite{liu2021diffsinger} proposes a conditional denoising diffusion probabilistic model (DDPM) and generates high-fidelity samples via iterative refinement. These methods aim to facilitate the first-stage neural network (a.k.a. acoustic model) of generating intermediate features. In contrast, our work focuses on neural singing voice vocoding in the second stage, which is relatively overlooked.

\subsection{Neural Vocoder}
Current vocoders can be categorized into several distinct families: 1) autoregressive models: WaveNet~\cite{oord2016wavenet} and WaveRNN~\cite{kalchbrenner2018efficient} generate waveforms autoregressively using dilated convolution layers, while the slow inference speed hinders their applications in high temporal resolution audio; 2) flow-based models~\cite{oord2018parallel,prenger2019waveglow} are proposed to address the above issues by transforming noise sequences of the same size in parallel, which utilize modern parallel computing processors to speed up sampling; 3) GAN-based models: MelGAN~\cite{kumar2019melgan} is the first work that successfully trains GANs for raw audio generation. Parallel WaveGAN~\cite{yamamoto2020parallel} uses a generator similar to WaveNet in structure and introduces multi-resolution STFT loss and adversarial loss, bringing improvement to waveform reconstruction. HIFI-GAN~\cite{kong2020hifigan} is proposed to model the periodic patterns matters and achieve both higher computational efficiency and sample quality than AR or flow-based models; and 4) most recently, researchers have proposed diffusion neural vocoders~\cite{chen2020wavegrad,kong2020diffwave} for waveform generation, which are built on prior work of score matching and diffusion probabilistic models. It is worth mentioning that~\citet{huang2021multisinger} does not implicitly focus upon the characteristics of the singing voices but aims to alleviate the model generalization in the multi-singer scenario. Different from previous works, we analyze the characteristics of singing voices and design a novel network for high-fidelity singing voice vocoding.

\section{Preliminary Analyses on Speech Vocoders} \label{analysis}
Singing voice is different from human speech for its long continuous pronunciation, high sampling rate and strong expressiveness. In this section, we revisit and conduct preliminary experiments to analyze how previous speech vocoders perform on the singing voice synthesis task. Specifically, we describe the glitches and high-frequency reconstruction challenges in detail, which motivate researchers to explicitly develop advanced models in singing voice vocoding.

\subsection{Glitches in Samples}
\begin{figure}[h]
  \centering
  \vspace{-5mm}
  \includegraphics[width=0.45\textwidth]{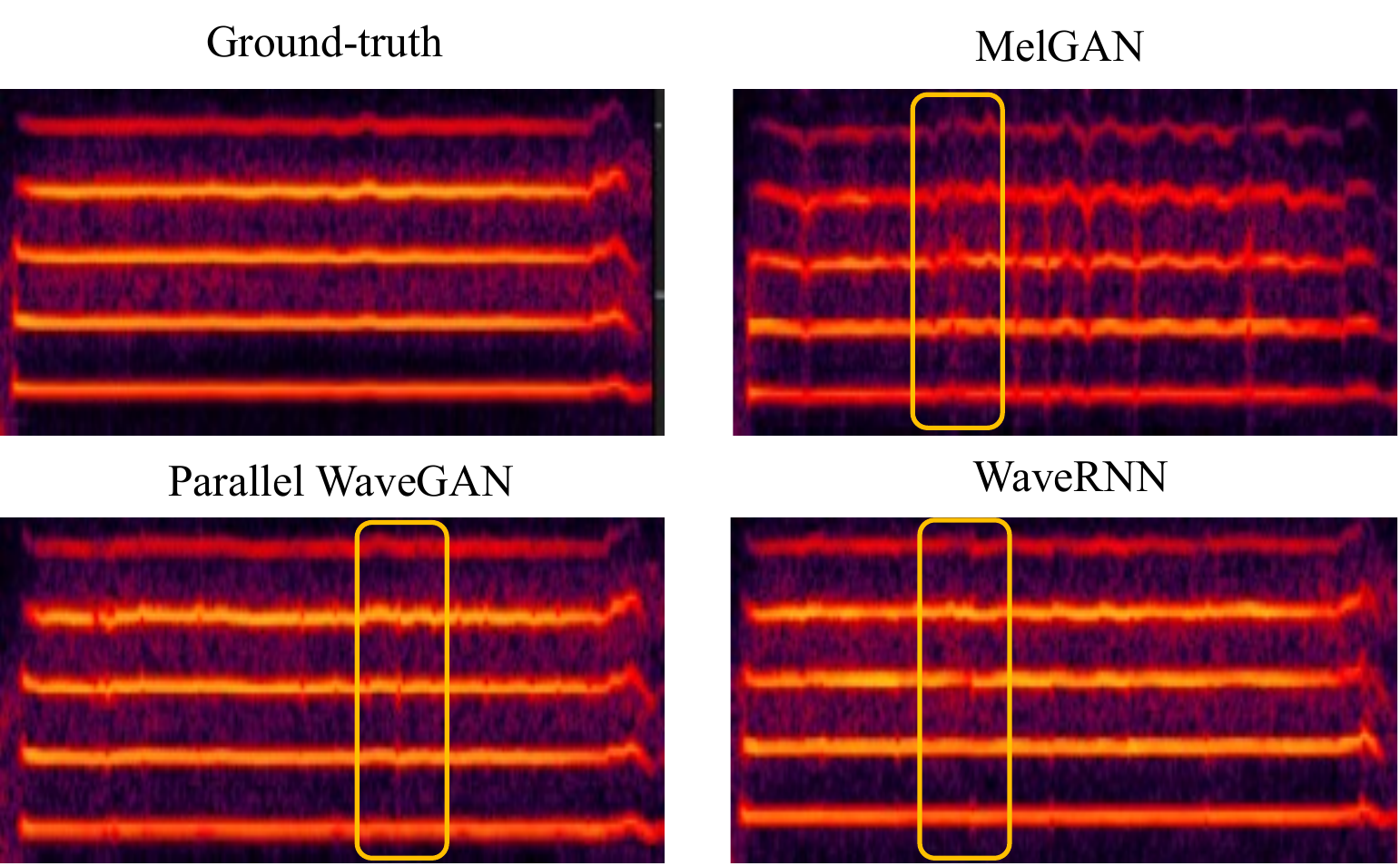}
  \caption{Visualization of glitch issues in generated samples.}
  \label{fig:Glitch}
   \vspace{-4mm}
\end{figure}

It is well-known~\cite{oord2016wavenet,kalchbrenner2018efficient} that the waveform segments sampled on each mel-spectrogram frame are not continuous, and thus neural vocoders usually require a sizeable receptive field to learn this continuity. In contrast to speeches, singing voices have even longer continuous pronunciation with smooth harmonics in the low-frequency parts, which brings difficulties in capturing the time dependencies of waveforms. As shown in Figure~\ref{fig:Glitch}, glitches could be easily observed in samples generated by previous speech vocoders (e.g., MelGAN~\cite{kumar2019melgan}, Parallel WaveGAN~\cite{yamamoto2020parallel}, and WaveRNN~\cite{kalchbrenner2018efficient}), especially when the duration of the singing sample exceeds limited receptive fields. 

\subsection{High-Frequency Reconstruction}

\begin{figure}[h]
\centering
 \vspace{-3mm}
\subfigure[Hissing noise in WaveRNN.]
{
  \label{fig:hissing}
  \includegraphics[width=0.45\textwidth]{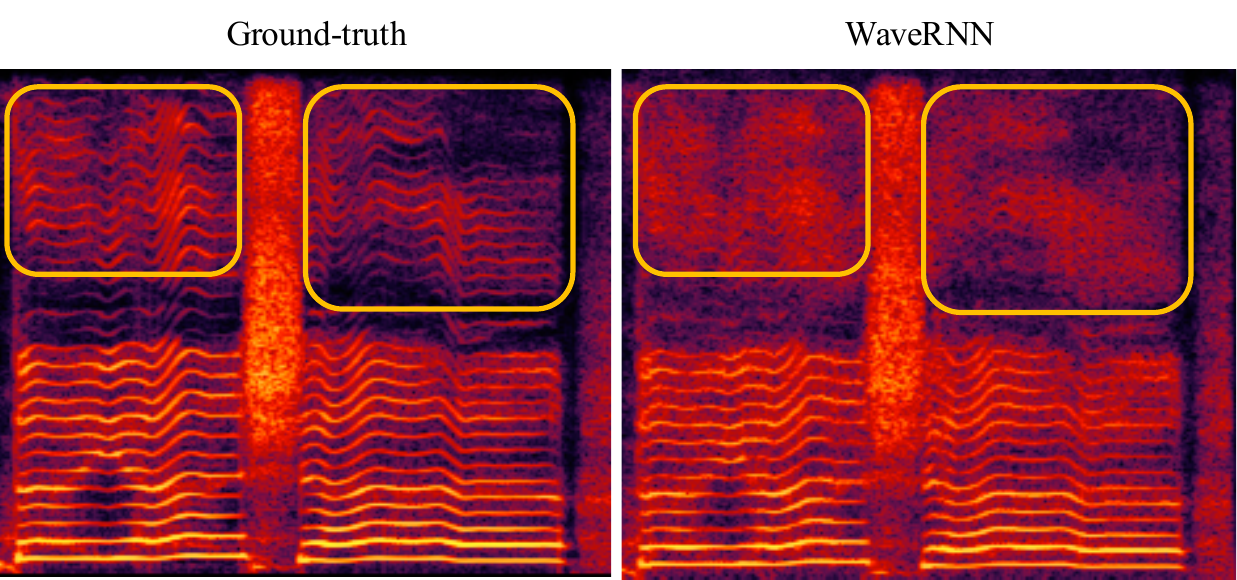}
}
\vspace{-2mm}
\subfigure[Metallic noise in Parallel WaveGAN.]
{
  \label{fig:metallic}
  \includegraphics[width=0.45\textwidth]{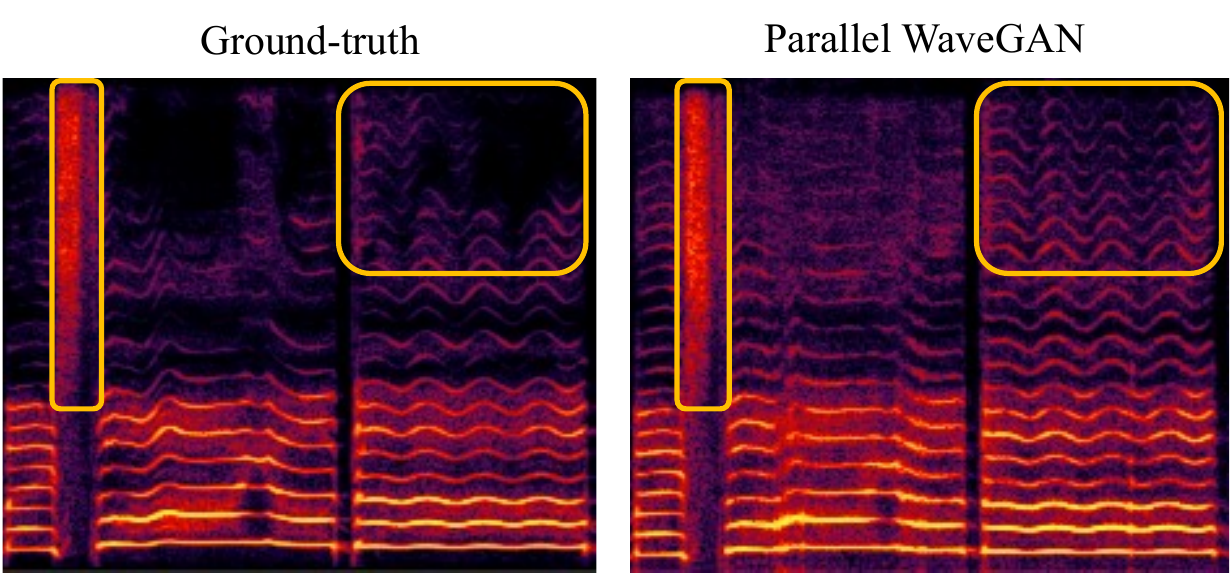}
}
\caption{Visualization of high-frequency reconstruction challenges.}
\label{fig:high} 
\vspace{-5mm}
\end{figure}

In contrast to the speech samples (e.g., 16k or 22kHz) in most scenarios~\cite{chen2021adaspeech,min2021meta}, singing voice recordings usually have a higher sampling rate (e.g., 24k or 48kHz) and contain rich high-frequency parts for improving expressiveness. It is challenging for neural vocoders to catch the short-term fluctuation in high-frequency signals, which in practice leads to distinct degradation of perceptual quality. 

As illustrated in Figure~\ref{fig:high}, we then visualize the mel-spectrograms generated by speech vocoders given the same mel-spectrogram sequence and have following observations: The autoregressive speech vocoder WaveRNN~\cite{kalchbrenner2018efficient} loses the attention to short-term fluctuation so that the apparent hissing noises in high-frequency bands are observed. The GAN-based model Parallel WaveGAN reconstructs the high-frequency parts simply by adding the metallic noises, far from meeting our expectations.


In summary, due to the long continuous pronunciation, high sampling rate, and strong expressiveness of singing voices, directly applying speech vocoder to singing voice generation results in glitches and poor high-frequency reconstruction. As such, designing a neural singing voice vocoder to explicitly alleviate these issues is an urgent demand.

\begin{figure*}[ht]
    \centering
    \vspace{-4mm}
    \subfigure[Generator]
    {
    \label{fig:Generator}
    \includegraphics[width=.6\textwidth]{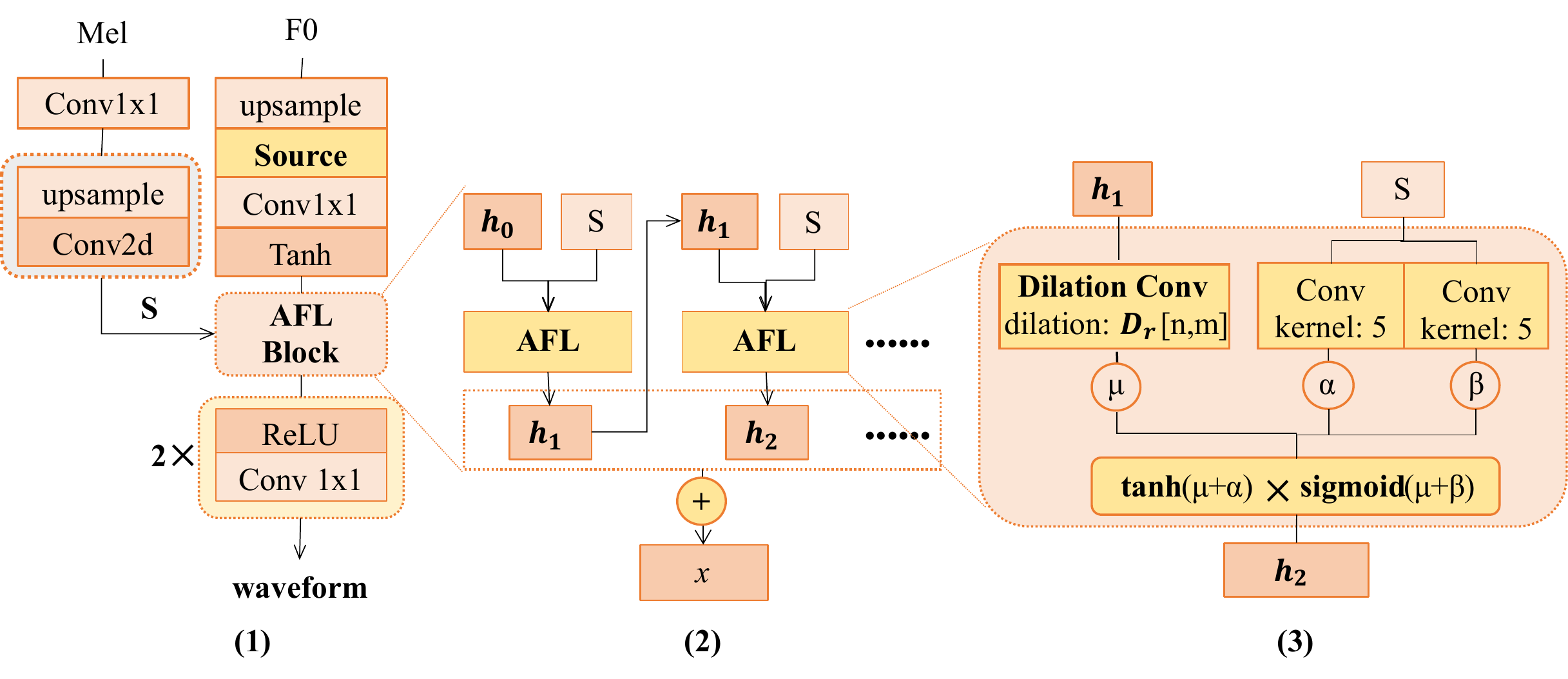}
    }
    \subfigure[Discriminator]
    {
    \label{fig:Discriminator}
    \includegraphics[width=0.35\textwidth]{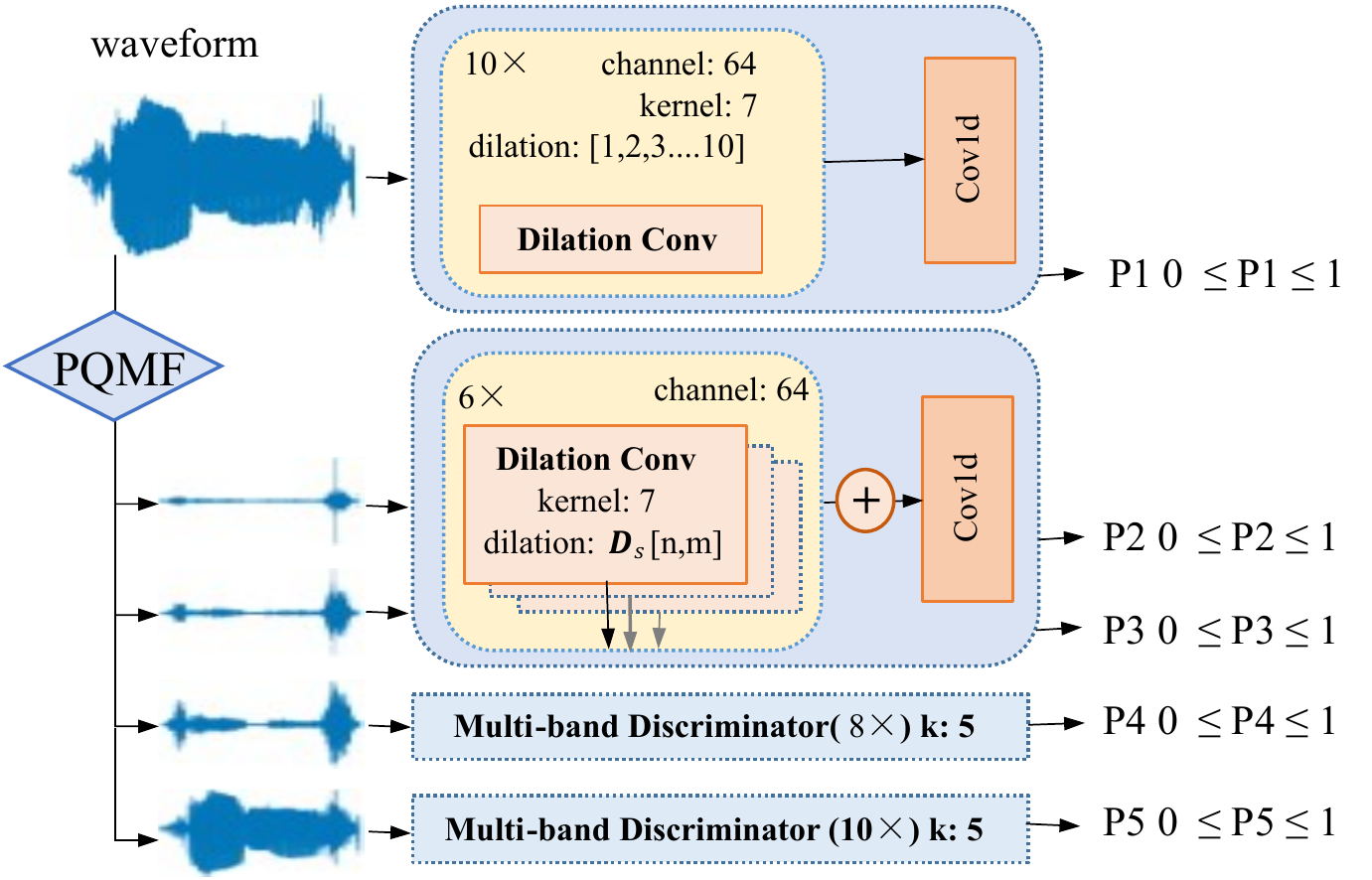}
    }
\vspace{-2mm}
\caption{The overall architecture of SingGAN. In subfigure(a), F0 denotes the fundamental frequency input. Source and AFL denote the source excitation and adaptive feature learning (AFL) filter, respectively. In subfigure(b), PQMF denotes the differentiable Pseudo Quadrature Mirror Filter bank to split singing voices into multiple frequency bands. $k$ denotes the kernel size.}
\vspace{-4mm}
\label{fig:architecture} 
\end{figure*}

\section{SingGAN}
This section presents SingGAN, a generative adversarial network designed for high-fidelity singing voice synthesis. The generative adversarial network jointly trains a powerful generator G, and a convolutional neural network (CNN) discriminator D, to synthesize time-domain waveforms given the corresponding mel-spectrogram. Firstly, we overview the motivation and overall designs, following which we describe the generator, discriminator and training objectives in detail. We illustrate SingGAN in Figure~\ref{fig:architecture}, and put more details of model architecture in Appendix~\ref{architecture} in the supplementary materials.

\subsection{Motivation}

It is well-known that the singing voices typically contain longer continuous pronunciation and rich high-frequency parts in the frequency domain for promoting expressiveness. Through the preliminary analyses in Section~\ref{analysis}, we empirically find that when directly applying speech models in singing voice vocoding, an apparent degradation in perceptual quality is observed due to the glitch issues and poor high-frequency reconstruction. 

In SingGAN, we propose several key techniques to complement the above issues: 1) to reduce glitches in the generated samples, SingGAN introduces a source-excitation generator driven by F0 (fundamental frequency) with the adaptive feature learning filters to expand receptive fields and stabilize long continuous signal generation; 2) SingGAN adopts the global discriminator to rebuild overall characteristics (e.g., singer identity and prosody). Further, it utilizes local multi-band discriminators to enrich low-frequency details and promote high-frequency reconstruction, and 3) to improve the training efficiency and the fidelity of the generated audio, SingGAN utilizes the sub-band feature matching penalty loss and auxiliary spectrogram losses as the joint training approach. 

\subsection{Generator}
Figure~\ref{fig:Generator} depicts the architecture of the generator with two vital components: a source module that produces a sine-based excitation signal and a filter module that uses the dilated convolutional network to convert the excitation into waveform. The generator expands receptive fields to capture time-dependencies in the long continuous pronunciations, preventing glitches from appearing in generated audio.

\textbf{Source excitation} 
The source excitation module generates gaussian noise in the unvoiced part and a sine-based excitation signal in the harmonic part. Specifically, 1) in the voiced segments, the excitation signal is a mixture of sine waveforms whose frequency values are determined by the fundamental frequency (f0) and its harmonics, and 2) in the unvoiced regions, the excitation signal is a sequence of gaussian noise. 

Given f0 sequences $\tilde{f}_{1: \tau}=\{\tilde{f}_{1}, \tilde{f}_{2}, \ldots \tilde{f}_{\tau}\}$, we get $f_{1: T}=\{f_{1}, f_{2} \ldots f_{T}\}$ by linear interpolation of $\tilde{f}_{1: \tau}$ to the same length $T$ of waveform. A sinusoidal excitation $e_{i j}$ that carries the $i$-th harmonic can be generated:
\begin{equation}
      e_{i j}=\left\{\begin{array}{cc}
      \sin (2 \pi \sum\limits_{j=1}^{T}\frac{i f_{j}}{S r} +\phi_{i}) & \text { if } f_{j}>0 \\
      Z_{\text {noise }} & \text { if } f_{j}=0
    \end{array}\right.,
\end{equation}

where $i \in\{1,2, \cdots W\}$, $j \in\{1,2, \cdots T\}$, we use $S_r$ and $W$ to denote the audio sampling rate and total number of harmonics, respectively. $e_{i j}$ denotes the excitation value of the $i$-th harmonic in $j$-th frame. $\phi_{i}$ is a phase number in the range of $[-\pi,\pi]$. In source excitation, frequency features would be transformed into $W$ initial harmonics matrix using the sine oscillator. In unvoiced regions where $f_{j}=0$, excitation would be replaced by gaussian noise $Z_{\text {noise }} \sim \mathcal{N}\left(0, \sigma^{2}\right)$.
    

\textbf{Adaptive feature learning filter} 
We propose an adaptive feature learning (AFL) filter block to convert the sinusoidal excitation $e_{i j}$ into output waveform $o_{i j}$. It is well-known that designing diverse receptive field patterns~\cite{oord2016wavenet,kalchbrenner2018efficient} is essential in catching time dependencies in waveforms. As such, the proposed AFL filter block utilizes couples of dilated convolutional layers to expand receptive field patterns and promote stability. 
As illustrated in Figure~\ref{fig:Generator}(2), each AFL filter block takes the sinusoidal excitation $e_{i j}$ and the upsampled mel features $c$ as input, generating the hidden state $h_{r}$ for the next block. In SingGAN, we design three AFL filter blocks in the generator, each containing 10 convolutional layers with dilation $D_{r}=\{1,2,4...512\}$. The channel size, kernel size, and window stride in each dilated convolutional layer are set to 64, 5, and 1, respectively. The gated activation unit (GAU)~\cite{oord2016wavenet} is added to each residual connection, which has been demonstrated for its efficiency in increasing the nonlinearity. In practice, the number of convolutional layers in each AFL filter block can be regulated to match one’s own requirement in a trade-off between synthesis speed and sample quality.


\subsection{Discriminator}
It is well-known that waveforms have fine-grained structures at different scales. The discriminator operating on downsampled audio will not have access to the high-frequency parts, and thus it is biased to conduct adversarial learning based on the low-frequency components only. To promote the naturalness of generated voice, these discriminators are supposed to learn discriminative features and output the possibilities in different frequency ranges. 

As illustrated in Figure~\ref{fig:Discriminator}, SingGAN adopts the global and local discriminators, which have identical network structures but operate on different scales. The global-level discriminator takes the full-band waveforms as input to capture and rebuild the overall characteristics of the singing voice (e.g., singer identity and prosody). After division by the differentiable Pseudo Quadrature Mirror Filter (PQMF) bank~\cite{yang2020multiband}, the four sub-band signals are respectively fed into the multiple local-level discriminators. These local discriminators output the possibilities at different scales, enriching the low-frequency details and facilitating high-frequency reconstruction.

In summary, the proposed global and local multi-level discriminators operate on waveforms at different scales and supervise the periodicity reconstruction in a wide range of frequency bands. 

\subsection{Training Loss}
The learning objectives should be carefully designed for stable training and faster convergence. In this section, we first describe the GAN adversarial loss. To improve the training efficiency and the fidelity of the generated audio, we adopt auxiliary multi-resolution STFT and mel-spectrogram loss as joint efforts for spectrogram supervision. Furthermore, we propose the sub-band feature matching penalty loss and decide our final objectives.

\subsubsection{Adversarial loss} 
For generative adversarial networks~\cite{goodfellow2014generative} which play competitive games, the generator is trained to fool the discriminator by updating the sample quality to be classified to a value almost equal to 1. The discriminator is trained to classify ground truth samples to 1, and the samples synthesized from the generator to 0. GAN losses for the generator $G$ and the discriminator $D$ are defined as:

\begin{equation}
    \nonumber
      L_{\text{adv}}\left(D_{k} ; G\right)=E_{x, s,f_{0}}\left[\left(1-D_{k}\left(x_{k}\right)\right)^{2}+\left(D_{k}\left(y_{k}\right)\right)^{2}\right],
    \end{equation}
  \begin{equation}
    \nonumber
      L_{\text{adv}}\left(G ; D_{k}\right)=E_{s, f_{0}}\left[\frac{1}{K} \sum_{k=1}^{K}\left(1-D_{k}\left(y_{k}\right)\right)^{2}\right],
    \end{equation}
    
where $x$ and $y$ denote the full-band ground truth and generated sample, respectively; $x_k$ or $y_k$ denotes the $k$-th sub-band waveforms. $s$ and $f_{0}$ denote the ground truth mel-spectrogram and fundamental frequency, respectively; $D_{k}$ represents $k$-th discriminators, and we set $K=5$ in SingGAN, including one full-band global discriminator and four sub-band local discriminators.

\subsubsection{Auxiliary Spectrogram losses} In addition to the GAN objective, auxiliary spectrogram losses have been demonstrated for their effectiveness in improving the training robustness and stability. Referring to the previous speech model~\cite{yamamoto2020parallel}, the multi-resolution STFT loss term pays much attention to the periodicity in waveform reconstruction. Besides, ~\citet{kong2020hifigan} has proved that time-frequency distribution could be captured effectively by jointly optimizing multi-resolution spectrogram loss and adversarial loss functions. 

\textbf{Multi-resolution STFT Loss} To stabilize the adversarial training process, SingGAN adopts a multi-resolution STFT (Short Time Fourier Transform) loss as follows:
\begin{equation}
  \nonumber
    L_{stft_{-} s c}(x, y)=\frac{\|\operatorname{STFT}(x)-\operatorname{STFT}(y)\|_{F}}{\|\operatorname{STFT}(x)\|_{F}},
\end{equation}
\begin{equation}
      \nonumber
      L_{stft_{-} mag}(x, y)=\frac{1}{N}\|\log (\operatorname{STFT}(x))-\log (\operatorname{STFT}(y))\|_{1},
\end{equation}
where $x$ and $y$ denote the ground-truth and predicted sample; $\|\cdot\|_{F}$ and $\|\cdot\|_{1}$ denote the Frobenius and L1 norms. We use $S T F T(\cdot)$ to denotes the function that transforms a waveform into the corresponding STFT spectrogram. $N$ denotes the number of elements in the magnitude; $L_{stft_{-}sc}$ and $L_{stft_{-} mag}$ denote the spectral convergence and log STFT magnitude, respectively.

The final multi-resolution STFT loss is the sum of $M_1$ losses with different analysis parameters(i.e., FFT size, window size, and hop size), and we set $M_1=3$ in SingGAN:
\begin{equation}
\nonumber
L_{s t f t}(x, y)=\frac{1}{M_1} \sum_{m=1}^{M_1}\left(L_{stft_{-} s c}^{(m)}(x, y)+L_{stft_{-} m a g}^{(m)}(x, y)\right).
\end{equation}

\textbf{Multi-resolution mel-spectrogram loss} To promote the training efficiency, SingGAN includes the multi-resolution mel-spectrogram loss. We define the multi-resolution mel-spectrogram loss as follows:
\begin{equation}
  \nonumber
    L_{mel_{-} s c}(x, y)=\frac{\|M E L(x)-M E L(y)\|_{F}}{\|M E L(x)\|_{F}},
    \end{equation}
  
\begin{equation}
  \nonumber
  L_{mel_{-} m a g}(x, y)=\frac{1}{N}\|\log (\operatorname{MEL}(x))-\log (\operatorname{MEL}(y))\|_{1},
\end{equation}

where $M E L(\cdot)$ denotes the function that transforms a waveform into the corresponding mel-spectrogram, and $N$ denotes the number of elements in the magnitude.
  
The final multi-resolution mel-spectrogram loss is the sum of $M_2$ losses with different analysis parameters, and we set $M_2=2$ in SingGAN:
\begin{equation}
  \nonumber
  L_{m e l}(x, y)=\frac{1}{M_2} \sum_{m=1}^{M_2}\left(L_{mel_{-} s c}^{(m)}(x, y)+L_{mel_{-} m a g}^{(m)}(x, y)\right).
\end{equation}

Our auxiliary spectrogram losses are defined as a linear combination of the multi-resolution STFT loss and the multi-resolution mel-spectrogram loss as follows:
\begin{equation}
\nonumber
  L_{a u x}(G)= L_{m e l}+\lambda L_{s t f t},
\end{equation}
where $L_{s t f t}$ and $L_{m e l}$ denote the multi-resolution STFT loss and mel-spectrogram loss, respectively. $\lambda$ denotes the trade-off between these two loss terms, and we set $\lambda=0.5$ in SingGAN.

\subsubsection{Sub-band feature matching penalty}
In addition to the GAN objectives and auxiliary spectrogram losses, we empirically find it helpful to penalize the mismatch of feature maps between the reference and generated samples in different frequency bands. Every intermediate feature of the discriminator is extracted, and the L1 distance between the ultimate output map of global and local discriminators in different scales is calculated. The sub-band feature matching penalty loss is defined as: 
\begin{equation}
\nonumber
    L_{f m}\left(G ; D_{k}\right)=E_{\left(x, s, f_{0}\right)}\left[\sum_{k=2}^{K}\left\|D_{k}\left(x_{k}\right)-D_{k}\left(y_{k}\right)\right\|_{1}\right].
\end{equation}

\subsubsection{Final loss}
Our final loss function in training the generator is defined as a linear combination of the adversarial loss, auxiliary spectrogram losses, and sub-band feature matching penalty loss. The final objectives for the generator and discriminator are presented as follows:
 
\begin{equation}
  \nonumber
    L_{D}= \sum_{k=1}^{K} L_{\text{adv}}\left(D_{k} ; G\right),
  \end{equation}
\begin{equation}
    \nonumber
      L_{G}=L_{a u x}(G) + \sum_{k=1}^{K}[\lambda_{1} L_{a d v}\left(G ; D_{k}\right)+\lambda_{2} L_{f m}\left(G ; D_{k}\right)],
\end{equation}
where $\lambda_{1}$ and $\lambda_{2} $ denote the hyperparameters in balancing loss terms, and we set $\lambda_{1}=4$ and $\lambda_{2}=10$ in our paper. $K$ denotes the number of discriminators, and we set $K=5$ in our paper. SingGAN has demonstrated its effectiveness and robustness in high-fidelity singing voice vocoding by jointly optimizing the loss functions mentioned above.

\section{Singing Voice Synthesis System}
\begin{figure}[h]
  \centering
  \includegraphics[width=0.5\textwidth]{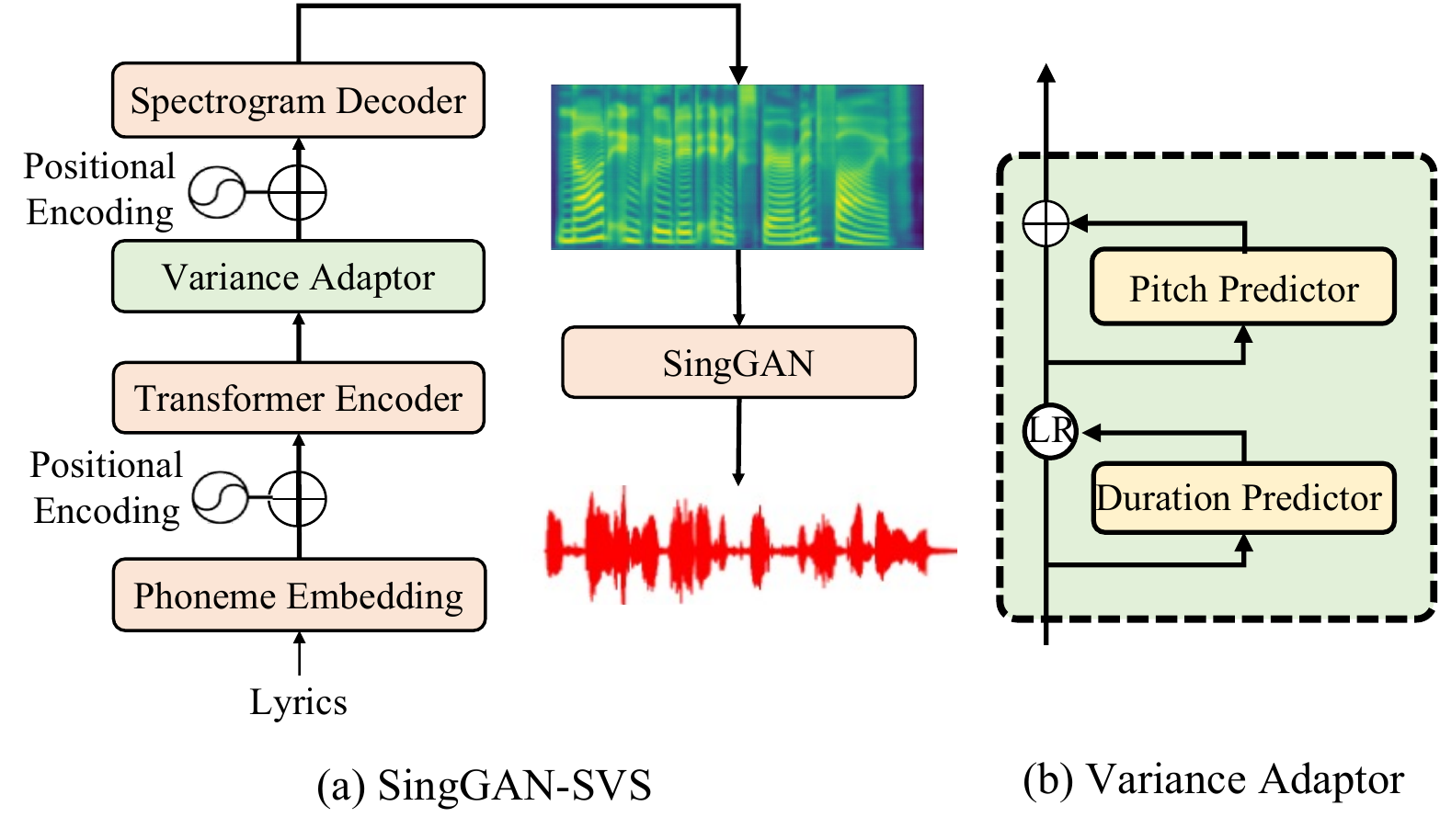}
  \vspace{-6mm}
  \caption{The overall architecture of SingGAN-SVS systems. In subfigure (a), we use the sinusoidal-like symbol to denote the positional encoding. In subfigure (b), we use "LR" to denote the length regulator proposed in FastSpeech~\cite{Fastspeech}.}
  \label{fig:appendix_arch}
  \vspace{-4mm}
  \end{figure}

To verify the effectiveness and robustness of the proposed SingGAN in the overall singing voice pipeline, we extend SingGAN to SingGAN-SVS for two-stage singing voice generation, including 1) a lyric-to-spectrogram generation module (a.k.a. acoustic model) to generate prosodic attributes with variance information; and 2) a conditional waveform generation module (i.e., SingGAN) to add the phase information and synthesize a detailed waveform. The overall architecture of SingGAN-SVS is illustrated in Figure~\ref{fig:appendix_arch}. 

\subsection{Lyric-to-spectrogram generation}
The architecture design of the lyric-to-spectrogram generation model in SingGAN-SVS refers to a non-autoregressive text-to-speech model FastSpeech 2~\cite{ren2020fastspeech} as the backbone. The transformer encoder first converts the lyrics sequence into the hidden representations. Then, the hidden representations would be expanded to match the length of the desired waveform output. Given the length-regulated sequence, the variance adaptor adds pitch variations and then the mel-spectrogram decoder generates high-fidelity mel-spectrograms. 

The final loss term in training the lyric-to-spectrogram generation model consists of the following parts: 1) mel reconstruction loss $\mathcal{L_{\text{mel}}}$: MSE between the predicted and ground-truth mel-spectrograms; 2) duration prediction loss $\mathcal{L_{\text{dur}}}$: MSE between the predicted and the ground-truth phoneme-level duration in log scale; and 3) pitch reconstruction loss $\mathcal{L_{\text{p}}}$: MSE between the ground-truth and predicted pitch spectrogram by pitch predictor.

\subsection{Spectrogram-to-waveform generation}
In SingGAN-SVS, we adopt SingGAN for a high-fidelity spectrogram-to-waveform generation. Due to the high sampling rate of singing voice audio (i.e., 24,000 samples per second) and the limited GPU memory, taking audio corresponding to the entire lyric sequence as input during training could be difficult. Therefore, we sample a small segment to synthesize the waveform before passing it to the SingGAN model.

\section{Experimental}
\subsection{Experimental Setup}
\paragraph{\textbf{Dataset}}
We use the internal singing voice dataset for fair and reproducible comparison with other models. The dataset contains Mandarin pop songs collected from 93 singers in a professional recording studio. All songs are saved in wav format, sampled at 24 kHz, and quantized by 16 bits. We randomly choose 340 utterances for validation and 60 utterances from 6 singers to be the test set. Additionally, to evaluate the generalization and robustness of SingGAN, we prepare utterances from unseen singers as the unseen test set. The unseen test set contains 5 utterances from each singer, including five males and five females. Following the common practice, we conduct preprocessing on the data: 1) we extract fundamental frequency (F0) from the waveforms using the public Parselmouth tool\footnote{\url{https://github.com/YannickJadoul/Parselmouth}}; and 2) we use 80-band mel-spectrograms as input conditions, where the FFT size, window stride, and hop size are set to 512, 512, and 128, respectively. 

\paragraph{\textbf{Model Configurations}}
The generator in SingGAN takes both log mel-spectrogram and F0 as input. To match the hop size, the upsampling rates of mel-spectrogram and F0 are set to $[8, 4, 4]$, 128, respectively. To smooth the low-frequency harmonics and reduce glitches in the continuous pronunciation, we use 8 harmonics in the source excitation. For discriminators, we set the number of convolutional layers, kernel size and dilated factor $D_{s}$ in high, middle and low frequency bands as 6, 7, \{1,1,1...1\}; 8, 5, \{1,2,3...8\} and 10, 5, \{1,2,3...10\}, respectively. The lyric-to-spectrogram generation model in SingGAN-SVS follows the basic configurations in FastSpeech 2~\cite{ren2020fastspeech}, which consists of 4 feed-forward transformer blocks in the phoneme encoder and mel-spectrogram decoder. More detailed information on architecture has been attached in Appendix~\ref{architecture}.

\begin{table*}[ht]
  \centering
  \scalebox{1.0}{
  \begin{tabular}{l|cc|cccc}
  \toprule
  \multirow{2}{*}{\bfseries Model} & \multicolumn{2}{c|}{\bfseries Sampling Speed} & \multicolumn{4}{c}{\bfseries Audio Quality}\\
  &\bfseries Parameters &\bfseries RTF ($\downarrow$) &\bfseries MOS ($\uparrow$) &\bfseries MCD ($\downarrow$) &\bfseries STOI ($\uparrow$)  &\bfseries PESQ ($\uparrow$)\\
  \midrule
  GT                                     &  /         & /          & 4.25$\pm$0.10    & /  &  /   &  / \\
  \midrule
  WaveRNN~\cite{kalchbrenner2018efficient}  & 4.48M      & 1.29      & 3.88$\pm$0.12  & 3.26 & 0.69  &  2.24\\
  NSF~\cite{wang2019neural}              & 1.2M       & 0.03       & 3.12$\pm$0.08    & 4.43 & 0.70  &  1.89\\
  MelGAN~\cite{kumar2019melgan}          & 5.31M      & 0.01       & 3.24$\pm$0.10    & 1.79 & 0.78  &  2.65\\
  Parallel WaveGAN~\cite{yamamoto2020parallel} & 1.44M   & 0.03    & 3.58$\pm$0.11    & 1.60 & 0.79  &  2.70\\
  HIFIGAN~\cite{kong2020hifigan}         &  13.92M      & 0.04     & 3.82$\pm$0.11    & 1.42 & 0.77  &  2.72\\
  Diffwave~\cite{kong2020diffwave}       &  2.62M      & 0.05      & 3.90$\pm$0.13    & 1.45 & 0.77  &  2.50 \\
  \midrule
  SingGAN                                & 1.59M      & 0.02     & \textbf{4.05$\pm$0.09}  & \textbf{1.14} & \textbf{0.80} & \textbf{2.75}\\
  \bottomrule
  \end{tabular}}
  \vspace{2mm}
  \caption{Comparison with other neural vocoders in term of quality and synthesis speed. The evaluation is conducted on a server with 1 NVIDIA 2080Ti GPU and batch size 1. For sampling, we use 6 iterations in Diffwave, following~\protect\cite{github2021DiffWaveVocoder}.}
  \vspace{-6mm}
  \label{table:mos1}
\end{table*}

\paragraph{\textbf{Training and Evaluation}}
At the training stage, the discriminators are turned on starting from 50k steps to warm up the generator. The adversarial loss is computed by the average of per-time step scalar predictions with multi-band discriminators. The auxiliary spectrogram losses are computed by the sum of different analysis parameters for STFT and mel-spectrogram inversion as described in Appendix~\ref{loss}. SingGAN is trained for 400k steps with a RAdam optimizer with $\beta_{1}=0.9, \beta_{2}=0.98, \epsilon=10^{-9}$ on 1 NVIDIA 2080Ti GPU, with the batch size of 64 sentences on each GPU. For baseline vocoders, we train them from scratch in the singing voice dataset. GitHub implementations are used for reproducibility, and the configurations follow their original papers. More information about objective and subjective evaluations has been attached in Appendix~\ref{appendix:evaluation}.


\vspace{-3mm}
\subsection{Comparison with other models} \label{comparison}
We compare our SingGAN in terms of audio quality and sampling speed with competing baselines, including 1) GT, the ground truth audio; 2) WaveRNN~\cite{kalchbrenner2018efficient}, which autoregressively generates audio with a recurrent neural network; 3) NSF~\cite{wang2019neural}, the statistical parametric speech synthesis model with neural source-filter. 4) MelGAN~\cite{kumar2019melgan}, Parallel WaveGAN~\cite{yamamoto2020parallel} and HIFI-GAN V1~\cite{kong2020hifigan}, three popular GAN-based models for fast and high-quality audio synthesis. 5) Diffwave~\cite{kong2020diffwave}, the recently proposed diffusion probabilistic model for speech synthesis, and we use 6 iterations during decoding following the original implementation~\cite{kong2020diffwave}. For easy comparison, the results are compiled and presented in Table~\ref{table:mos1}, and we have the following observations:

In terms of audio quality, SingGAN achieves the highest MOS with a gap of 0.09 compared to the ground-truth audio, which indicates that the synthesized audio is nearly indistinguishable from the human voices. SingGAN outperforms the best publicly available models, successfully reducing glitches issues and boosting high-frequency reconstruction in singing voice vocoding. For objective evaluation, SingGAN also demonstrates a significant improvement in MCD, STOI, and PESQ. In terms of inference speed, SingGAN further enjoys computational efficiency competitive to baseline architectures, enabling a sampling speed of 50x faster than real-time on a single NVIDIA 2080Ti without engineered kernels.

\subsection{Ablation Studies} \label{ablation}
We conduct ablation studies to demonstrate the effectiveness of several components in SingGAN, including the: 1) source excitation; 2) adaptive feature learning filter; 3) global and local discriminators, and 4) loss objectives, including auxiliary spectrogram loss and sub-band feature matching penalty loss. As illustrated in Table~\ref{table:mos2}, we conduct CMOS (comparative mean opinion score) evaluations and have the following findings:

\begin{figure}[h]
  \centering
   \vspace{-3mm}
  \includegraphics[width=0.42\textwidth]{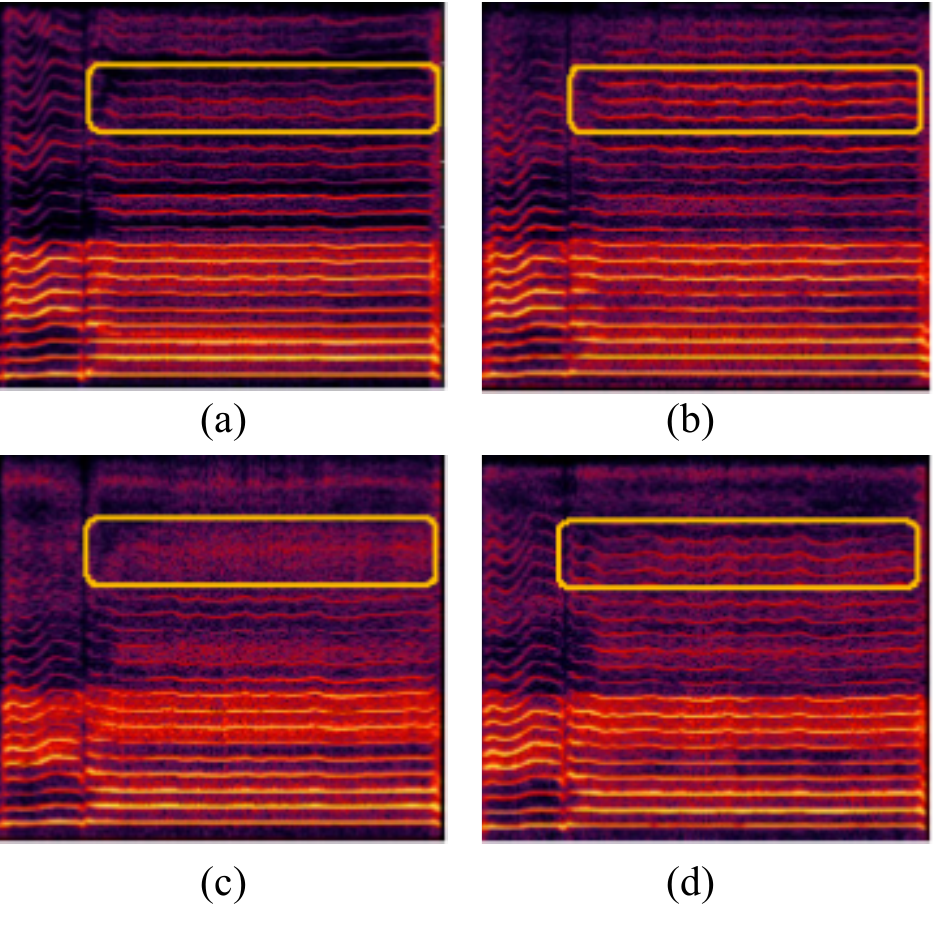}
 \caption{Visualization of the generated samples with varying auxiliary spectrogram losses. (a) Ground-truth. (b) With multi-resolution STFT loss only. The brighter spectrogram with false energy reconstruction is observed. (c) With multi-resolution mel-spectrogram loss only. The over-smooth spectrogram in high-frequency parts is observed. (d) Our joint method in auxiliary spectrogram losses.}
  \label{fig:Mel_Loss}
   \vspace{-3mm}
\end{figure}

Removing the source excitation eliminates the model's capacity to capture time dependencies in long continuous voice, and a distinct degradation in perceptual quality is observed. Replacing the AFL filter by dilated convolutions shows a relatively small but noticeable degradation in both sample quality and sampling speed. Removing the multi-level discriminators has decreased audio quality, demonstrating their efficiency in enriching low-frequency details and boosting high-frequency reconstruction. Also, removing the sub-band feature matching penalty loss results in the lower perceptual quality and slower convergence of GAN-based models.

In addition, we conduct ablation studies to verify the effectiveness of auxiliary spectrogram supervision, including the multi-resolution mel-spectrogram and multi-resolution STFT loss. We illustrate the spectrogram of the generated samples in Figure~\ref{fig:Mel_Loss}, and empirically find that arbitrary absence of these two loss terms results in worse predictions: Removing the mel-spectrogram loss prevents models from proper energy reconstruction, leading to the brighter spectrogram. Also, removing the STFT loss causes distinct over-smooth predictions in the high-frequency parts. 

\begin{table}[ht]
  \centering
   \vspace{-2mm}
    \begin{tabular}{lccccc}
      \toprule
       \bfseries Model  & \bfseries RTF ($\downarrow$) & \bfseries CMOS ($\uparrow$) \\
       \midrule
       SingGAN                &0.02 & 0.00   \\
       \midrule
       \quad\quad - source excitation   &0.02 & -0.28  \\
       \quad\quad - adaptive feature learning filter       &0.04 & -0.10  \\
       \quad\quad - discriminators       &0.02 & -0.22  \\
       \quad\quad - auxiliary spectrogram loss        &0.02 & -0.16  \\
       \quad\quad - sub-band feature matching loss       &0.02 & -0.12  \\
       \bottomrule
       \end{tabular}
       \vspace{2mm}
       \caption{Ablation study results. Comparison of the effect of each component in terms of quality and synthesis speed.}
        \vspace{-4mm}
       \label{table:mos2}
      \end{table}

      \begin{table}[ht]
        \centering
        \vspace{-6mm}
      \small
        \begin{tabular}{lcccc}
          \toprule
          \bfseries Model    & \bfseries MOS ($\uparrow$) & \bfseries MCD ($\downarrow$) &  \bfseries STOI ($\uparrow$)  &  \bfseries PESQ ($\uparrow$) \\
          \midrule
          GT               &  4.36$\pm$0.11 & / &  / & / \\
          \midrule
          WaveRNN          & 3.92$\pm$0.14  & 1.49 & 0.66  & 2.67  \\
          NSF              & 3.44$\pm$0.10  & 2.46 & 0.64 &  1.84 \\
          MelGAN            & 3.62$\pm$0.08 & 1.40 & 0.68 &  2.72 \\
          ParallelWaveGAN  & 3.81$\pm$0.11 & 1.00 & 0.68 &  2.78 \\
          HIFIGAN          & 3.97$\pm$0.11  & 1.04 & 0.73 &  2.45  \\
          Diffwave         & 3.94$\pm$0.14  & 1.02 & 0.74 &  2.75 \\
          \midrule
          \bfseries SingGAN          & \textbf{4.08$\pm$0.09}  & \textbf{0.92} & \textbf{0.78} & \textbf{2.81}  \\
          \bottomrule
          \end{tabular}
          \vspace{2mm}
          \caption{Comparison with other neural vocoders of synthesized utterances for unseen singers.}
          \vspace{-6mm}
          \label{table:mos5}
        \end{table}

\subsection{Generalization to Unseen Singers}\label{Generalization}
To evaluate the generalization ability of SingGAN, we prepare utterances from unseen singers (including five males and five females) as the additional test set, which contains 5 utterances from each singer. In an attempt to provide an easily comparable metric to evaluate this generalization, we presents the experimental results for the mel-spectrogram inversion of the unseen singers in Table~\ref{table:mos5}. Interestingly, we notice that the SingGAN scores $4.08$ and achieves state-of-the-art results for out-of-domain generalization, with a gap of $0.28$ compared to the ground truth audio. This experiment verifies that SingGAN is able to learn a singer-invariant mapping of mel-spectrograms to raw waveforms, demonstrating its generalizability to entirely new (unseen) singers outside the train set. Additionally, the tendency of difference in MOS scores of the proposed models is similar with the result shown in Section~\ref{comparison}, which exhibits generalization across different datasets.

\subsection{End-to-End Singing Voice Synthesis} 

We conduct an additional experiment to examine the effectiveness of the proposed model when extended to an end-to-end singing voice synthesis system SingGAN-SVS. We conduct MOS evaluation and compare with several systems, including 1) GT, where we first convert the ground truth audio into mel-spectrograms, and then convert the mel-spectrograms back to audio using SingGAN; 2) FastSpeech 2~\cite{ren2020fastspeech} + HIFI-GAN~\cite{kong2020hifigan}, we first utilize the acoustic model to generate spectrograms, and then transform them into waveforms using HIFI-GAN; and 3) FastSpeech 2s~\cite{ren2020fastspeech}, the end-to-end text-to-waveform generation model. The MOS results have been presented in Table~\ref{table:mos4}, and we have the following observations: SingGAN-SVS achieves the highest MOS with a gap of $0.11$ compared to the ground truth audio, which indicates that the synthesized audio is nearly indistinguishable from the human voices. SingGAN-SVS outperforms all publicly available baselines, demonstrating the efficiency and robustness of SingGAN in the overall end-to-end singing voice synthesis pipeline. For objective evaluation, ProDiff also respectively demonstrates a significant improvement of $2.08$, $0.43$, and $1.1$ in MCD, PESQ, and STOI, compared to the best baseline system.

\begin{table}[h]
  \centering
   \vspace{-2mm}
  \begin{tabular}{lcccc}
  \toprule
  \bfseries  Model            & \bfseries MOS  ($\uparrow$) &\bfseries MCD ($\downarrow$) &  \bfseries STOI ($\uparrow$)  &  \bfseries PESQ ($\uparrow$)  \\
  \midrule
  GT                &   4.05$\pm$0.09         &  / &  / & / \\
  \midrule
  FastSpeech 2            &   3.51$\pm$0.07         & 5.90	&0.17 	&1.32  \\
  FastSpeech 2s           &   3.40$\pm$0.07         & 5.56	&0.16 	&1.24 \\  
  SingGAN-SVS             & \bfseries 3.84$\pm$0.08 & \bfseries 3.48	& \bfseries 0.59   & \bfseries 2.34    \\
  \bottomrule
\end{tabular}
\vspace{2mm}
\caption{Comparison with other singing voice synthesis systems.}
\vspace{-6mm}
\label{table:mos4}
\end{table}

\vspace{-4mm}
\section{Conclusion}
We proposed SingGAN, a novel generative adversarial network designed for high-fidelity singing voice vocoding. Considering the long continuous pronunciation, high sampling rate, and strong expressiveness of the singing voices, we analyzed and revisited the glitches and poor high-frequency reconstruction when adopting previous speech models in singing voice vocoding. SingGAN employed the source excitation with adaptive feature learning filters to expand receptive fields and stabilize long continuous signal generation, efficiently alleviating the generated samples' glitch problem. To enrich low-frequency details and promote high-frequency reconstruction, SingGAN introduces competitive games with global and local discriminators at different scales. The learning objectives had been carefully designed for stable training and faster convergence. With auxiliary spectrogram losses and sub-band feature matching penalty loss, the training efficiency and the fidelity of the generated audio had been promoted.
Experimental results demonstrated that SingGAN outperformed the best publicly available neural vocoder for audio quality, even comparable to the human level. In terms of synthesis speed, SingGAN enables a sample speed of 50x faster than real-time on a single NVIDIA 2080Ti GPU. We further show that SingGAN generalized well to the mel-spectrogram inversion of unseen singers, and the extensive model SingGAN-SVS further enjoyed a two-stage pipeline to transform the music scores into singing voices. We envisage that our work could serve as a basis for future singing voice synthesis studies.

\section*{Acknowledgements}
This work was supported in part by the Zhejiang Natural Science Foundation LR19F020006 and National Key R\&D Program of China under Grant No.61836002 and No.62072397.

\bibliographystyle{ACM-Reference-Format}
\bibliography{sample-base}


\begin{thebibliography}{37}


\ifx \showCODEN    \undefined \def \showCODEN     #1{\unskip}     \fi
\ifx \showDOI      \undefined \def \showDOI       #1{#1}\fi
\ifx \showISBNx    \undefined \def \showISBNx     #1{\unskip}     \fi
\ifx \showISBNxiii \undefined \def \showISBNxiii  #1{\unskip}     \fi
\ifx \showISSN     \undefined \def \showISSN      #1{\unskip}     \fi
\ifx \showLCCN     \undefined \def \showLCCN      #1{\unskip}     \fi
\ifx \shownote     \undefined \def \shownote      #1{#1}          \fi
\ifx \showarticletitle \undefined \def \showarticletitle #1{#1}   \fi
\ifx \showURL      \undefined \def \showURL       {\relax}        \fi
\providecommand\bibfield[2]{#2}
\providecommand\bibinfo[2]{#2}
\providecommand\natexlab[1]{#1}
\providecommand\showeprint[2][]{arXiv:#2}

\bibitem[\protect\citeauthoryear{Chen, Tan, Luan, Qin, and Liu}{Chen
  et~al\mbox{.}}{2020a}]%
        {chen2020hifisinger}
\bibfield{author}{\bibinfo{person}{Jiawei Chen}, \bibinfo{person}{Xu Tan},
  \bibinfo{person}{Jian Luan}, \bibinfo{person}{Tao Qin}, {and}
  \bibinfo{person}{Tie-Yan Liu}.} \bibinfo{year}{2020}\natexlab{a}.
\newblock \showarticletitle{HiFiSinger: Towards High-Fidelity Neural Singing
  Voice Synthesis}.
\newblock \bibinfo{journal}{\emph{arXiv preprint arXiv:2009.01776}}
  (\bibinfo{year}{2020}).
\newblock


\bibitem[\protect\citeauthoryear{Chen, Tan, Li, Liu, Qin, Zhao, and Liu}{Chen
  et~al\mbox{.}}{2021}]%
        {chen2021adaspeech}
\bibfield{author}{\bibinfo{person}{Mingjian Chen}, \bibinfo{person}{Xu Tan},
  \bibinfo{person}{Bohan Li}, \bibinfo{person}{Yanqing Liu},
  \bibinfo{person}{Tao Qin}, \bibinfo{person}{Sheng Zhao}, {and}
  \bibinfo{person}{Tie-Yan Liu}.} \bibinfo{year}{2021}\natexlab{}.
\newblock \showarticletitle{Adaspeech: Adaptive text to speech for custom
  voice}.
\newblock \bibinfo{journal}{\emph{arXiv preprint arXiv:2103.00993}}
  (\bibinfo{year}{2021}).
\newblock


\bibitem[\protect\citeauthoryear{Chen, Zhang, Zen, Weiss, Norouzi, and
  Chan}{Chen et~al\mbox{.}}{2020b}]%
        {chen2020wavegrad}
\bibfield{author}{\bibinfo{person}{Nanxin Chen}, \bibinfo{person}{Yu Zhang},
  \bibinfo{person}{Heiga Zen}, \bibinfo{person}{Ron~J Weiss},
  \bibinfo{person}{Mohammad Norouzi}, {and} \bibinfo{person}{William Chan}.}
  \bibinfo{year}{2020}\natexlab{b}.
\newblock \showarticletitle{WaveGrad: Estimating Gradients for Waveform
  Generation}.
\newblock  (\bibinfo{year}{2020}).
\newblock


\bibitem[\protect\citeauthoryear{Cui, Ren, Liu, Chen, Huang, Lei, and Zhao}{Cui
  et~al\mbox{.}}{2021}]%
        {cui2021emovie}
\bibfield{author}{\bibinfo{person}{Chenye Cui}, \bibinfo{person}{Yi Ren},
  \bibinfo{person}{Jinglin Liu}, \bibinfo{person}{Feiyang Chen},
  \bibinfo{person}{Rongjie Huang}, \bibinfo{person}{Ming Lei}, {and}
  \bibinfo{person}{Zhou Zhao}.} \bibinfo{year}{2021}\natexlab{}.
\newblock \showarticletitle{EMOVIE: A Mandarin Emotion Speech Dataset with a
  Simple Emotional Text-to-Speech Model}.
\newblock \bibinfo{journal}{\emph{arXiv preprint arXiv:2106.09317}}
  (\bibinfo{year}{2021}).
\newblock


\bibitem[\protect\citeauthoryear{Goodfellow, Pouget-Abadie, Mirza, Xu,
  Warde-Farley, Ozair, Courville, and Bengio}{Goodfellow et~al\mbox{.}}{2014}]%
        {goodfellow2014generative}
\bibfield{author}{\bibinfo{person}{Ian~J. Goodfellow}, \bibinfo{person}{Jean
  Pouget-Abadie}, \bibinfo{person}{Mehdi Mirza}, \bibinfo{person}{Bing Xu},
  \bibinfo{person}{David Warde-Farley}, \bibinfo{person}{Sherjil Ozair},
  \bibinfo{person}{Aaron Courville}, {and} \bibinfo{person}{Yoshua Bengio}.}
  \bibinfo{year}{2014}\natexlab{}.
\newblock \showarticletitle{Generative Adversarial Networks}.
\newblock  (\bibinfo{year}{2014}).
\newblock
\showeprint[arxiv]{1406.2661}~[stat.ML]


\bibitem[\protect\citeauthoryear{Gu, Yin, Rao, Wan, Tang, Zhang, Chen, Wang,
  and Ma}{Gu et~al\mbox{.}}{2020}]%
        {gu2020bytesing}
\bibfield{author}{\bibinfo{person}{Yu Gu}, \bibinfo{person}{Xiang Yin},
  \bibinfo{person}{Yonghui Rao}, \bibinfo{person}{Yuan Wan},
  \bibinfo{person}{Benlai Tang}, \bibinfo{person}{Yang Zhang},
  \bibinfo{person}{Jitong Chen}, \bibinfo{person}{Yuxuan Wang}, {and}
  \bibinfo{person}{Zejun Ma}.} \bibinfo{year}{2020}\natexlab{}.
\newblock \showarticletitle{ByteSing: A Chinese Singing Voice Synthesis System
  Using Duration Allocated Encoder-Decoder Acoustic Models and WaveRNN
  Vocoders}.
\newblock \bibinfo{journal}{\emph{arXiv preprint arXiv:2004.11012}}
  (\bibinfo{year}{2020}).
\newblock


\bibitem[\protect\citeauthoryear{Huang, Chen, Ren, Liu, Cui, and Zhao}{Huang
  et~al\mbox{.}}{2021}]%
        {huang2021multisinger}
\bibfield{author}{\bibinfo{person}{Rongjie Huang}, \bibinfo{person}{Feiyang
  Chen}, \bibinfo{person}{Yi Ren}, \bibinfo{person}{Jinglin Liu},
  \bibinfo{person}{Chenye Cui}, {and} \bibinfo{person}{Zhou Zhao}.}
  \bibinfo{year}{2021}\natexlab{}.
\newblock \showarticletitle{Multi-Singer: Fast Multi-Singer Singing Voice
  Vocoder With A Large Scale Corpus}.
\newblock  (\bibinfo{year}{2021}).
\newblock


\bibitem[\protect\citeauthoryear{Huang, Lam, Wang, Su, Yu, Ren, and Zhao}{Huang
  et~al\mbox{.}}{2022a}]%
        {huang2022fastdiff}
\bibfield{author}{\bibinfo{person}{Rongjie Huang}, \bibinfo{person}{Max~WY
  Lam}, \bibinfo{person}{Jun Wang}, \bibinfo{person}{Dan Su},
  \bibinfo{person}{Dong Yu}, \bibinfo{person}{Yi Ren}, {and}
  \bibinfo{person}{Zhou Zhao}.} \bibinfo{year}{2022}\natexlab{a}.
\newblock \showarticletitle{FastDiff: A Fast Conditional Diffusion Model for
  High-Quality Speech Synthesis}.
\newblock \bibinfo{journal}{\emph{arXiv preprint arXiv:2204.09934}}
  (\bibinfo{year}{2022}).
\newblock


\bibitem[\protect\citeauthoryear{Huang, Ren, Liu, Cui, and Zhao}{Huang
  et~al\mbox{.}}{2022b}]%
        {huang2022generspeech}
\bibfield{author}{\bibinfo{person}{Rongjie Huang}, \bibinfo{person}{Yi Ren},
  \bibinfo{person}{Jinglin Liu}, \bibinfo{person}{Chenye Cui}, {and}
  \bibinfo{person}{Zhou Zhao}.} \bibinfo{year}{2022}\natexlab{b}.
\newblock \showarticletitle{GenerSpeech: Towards Style Transfer for
  Generalizable Out-Of-Domain Text-to-Speech Synthesis}.
\newblock \bibinfo{journal}{\emph{arXiv preprint arXiv:2205.07211}}
  (\bibinfo{year}{2022}).
\newblock


\bibitem[\protect\citeauthoryear{Huang, Zhao, Liu, Liu, Ren, Zhang, and
  He}{Huang et~al\mbox{.}}{2022c}]%
        {huang2022transpeech}
\bibfield{author}{\bibinfo{person}{Rongjie Huang}, \bibinfo{person}{Zhou Zhao},
  \bibinfo{person}{Jinglin Liu}, \bibinfo{person}{Huadai Liu},
  \bibinfo{person}{Yi Ren}, \bibinfo{person}{Lichao Zhang}, {and}
  \bibinfo{person}{Jinzheng He}.} \bibinfo{year}{2022}\natexlab{c}.
\newblock \showarticletitle{TranSpeech: Speech-to-Speech Translation With
  Bilateral Perturbation}.
\newblock \bibinfo{journal}{\emph{arXiv preprint arXiv:2205.12523}}
  (\bibinfo{year}{2022}).
\newblock


\bibitem[\protect\citeauthoryear{Kalchbrenner, Elsen, Simonyan, Noury,
  Casagrande, Lockhart, Stimberg, Oord, Dieleman, and Kavukcuoglu}{Kalchbrenner
  et~al\mbox{.}}{2018}]%
        {kalchbrenner2018efficient}
\bibfield{author}{\bibinfo{person}{Nal Kalchbrenner}, \bibinfo{person}{Erich
  Elsen}, \bibinfo{person}{Karen Simonyan}, \bibinfo{person}{Seb Noury},
  \bibinfo{person}{Norman Casagrande}, \bibinfo{person}{Edward Lockhart},
  \bibinfo{person}{Florian Stimberg}, \bibinfo{person}{Aaron van~den Oord},
  \bibinfo{person}{Sander Dieleman}, {and} \bibinfo{person}{Koray
  Kavukcuoglu}.} \bibinfo{year}{2018}\natexlab{}.
\newblock \showarticletitle{Efficient neural audio synthesis}.
\newblock \bibinfo{journal}{\emph{arXiv preprint arXiv:1802.08435}}
  (\bibinfo{year}{2018}).
\newblock


\bibitem[\protect\citeauthoryear{Kim, Kim, Kong, and Yoon}{Kim
  et~al\mbox{.}}{2020}]%
        {kim2020glowtts}
\bibfield{author}{\bibinfo{person}{Jaehyeon Kim}, \bibinfo{person}{Sungwon
  Kim}, \bibinfo{person}{Jungil Kong}, {and} \bibinfo{person}{Sungroh Yoon}.}
  \bibinfo{year}{2020}\natexlab{}.
\newblock \showarticletitle{Glow-TTS: A Generative Flow for Text-to-Speech via
  Monotonic Alignment Search}.
\newblock  (\bibinfo{year}{2020}).
\newblock
\showeprint[arxiv]{2005.11129}~[eess.AS]


\bibitem[\protect\citeauthoryear{Kong, Kim, and Bae}{Kong
  et~al\mbox{.}}{2020a}]%
        {kong2020hifigan}
\bibfield{author}{\bibinfo{person}{Jungil Kong}, \bibinfo{person}{Jaehyeon
  Kim}, {and} \bibinfo{person}{Jaekyoung Bae}.}
  \bibinfo{year}{2020}\natexlab{a}.
\newblock \showarticletitle{HiFi-GAN: Generative Adversarial Networks for
  Efficient and High Fidelity Speech Synthesis}.
\newblock \bibinfo{journal}{\emph{arXiv preprint arXiv:2010.05646}}
  (\bibinfo{year}{2020}).
\newblock


\bibitem[\protect\citeauthoryear{Kong, Ping, Huang, Zhao, and Catanzaro}{Kong
  et~al\mbox{.}}{2020b}]%
        {kong2020diffwave}
\bibfield{author}{\bibinfo{person}{Zhifeng Kong}, \bibinfo{person}{Wei Ping},
  \bibinfo{person}{Jiaji Huang}, \bibinfo{person}{Kexin Zhao}, {and}
  \bibinfo{person}{Bryan Catanzaro}.} \bibinfo{year}{2020}\natexlab{b}.
\newblock \showarticletitle{DiffWave: A Versatile Diffusion Model for Audio
  Synthesis}.
\newblock  (\bibinfo{year}{2020}).
\newblock


\bibitem[\protect\citeauthoryear{Kubichek}{Kubichek}{1993}]%
        {kubichek1993mel}
\bibfield{author}{\bibinfo{person}{Robert Kubichek}.}
  \bibinfo{year}{1993}\natexlab{}.
\newblock \showarticletitle{Mel-cepstral distance measure for objective speech
  quality assessment}.
\newblock   \bibinfo{volume}{1} (\bibinfo{year}{1993}),
  \bibinfo{pages}{125--128}.
\newblock


\bibitem[\protect\citeauthoryear{Kumar, Kumar, de~Boissiere, Gestin, Teoh,
  Sotelo, de~Br{\'e}bisson, Bengio, and Courville}{Kumar et~al\mbox{.}}{2019}]%
        {kumar2019melgan}
\bibfield{author}{\bibinfo{person}{Kundan Kumar}, \bibinfo{person}{Rithesh
  Kumar}, \bibinfo{person}{Thibault de Boissiere}, \bibinfo{person}{Lucas
  Gestin}, \bibinfo{person}{Wei~Zhen Teoh}, \bibinfo{person}{Jose Sotelo},
  \bibinfo{person}{Alexandre de Br{\'e}bisson}, \bibinfo{person}{Yoshua
  Bengio}, {and} \bibinfo{person}{Aaron~C Courville}.}
  \bibinfo{year}{2019}\natexlab{}.
\newblock \showarticletitle{Melgan: Generative adversarial networks for
  conditional waveform synthesis}.
\newblock  (\bibinfo{year}{2019}), \bibinfo{pages}{14910--14921}.
\newblock


\bibitem[\protect\citeauthoryear{Li, Liu, Liu, Zhao, and Liu}{Li
  et~al\mbox{.}}{2019}]%
        {li2019neural}
\bibfield{author}{\bibinfo{person}{Naihan Li}, \bibinfo{person}{Shujie Liu},
  \bibinfo{person}{Yanqing Liu}, \bibinfo{person}{Sheng Zhao}, {and}
  \bibinfo{person}{Ming Liu}.} \bibinfo{year}{2019}\natexlab{}.
\newblock \showarticletitle{Neural speech synthesis with transformer network}.
\newblock  \bibinfo{volume}{33}, \bibinfo{number}{01} (\bibinfo{year}{2019}),
  \bibinfo{pages}{6706--6713}.
\newblock


\bibitem[\protect\citeauthoryear{Liu, Li, Ren, Chen, Liu, and Zhao}{Liu
  et~al\mbox{.}}{2021}]%
        {liu2021diffsinger}
\bibfield{author}{\bibinfo{person}{Jinglin Liu}, \bibinfo{person}{Chengxi Li},
  \bibinfo{person}{Yi Ren}, \bibinfo{person}{Feiyang Chen},
  \bibinfo{person}{Peng Liu}, {and} \bibinfo{person}{Zhou Zhao}.}
  \bibinfo{year}{2021}\natexlab{}.
\newblock \showarticletitle{Diffsinger: Singing voice synthesis via shallow
  diffusion mechanism}.
\newblock \bibinfo{journal}{\emph{arXiv preprint arXiv:2105.02446}}
  \bibinfo{volume}{2} (\bibinfo{year}{2021}).
\newblock


\bibitem[\protect\citeauthoryear{Min, Lee, Yang, and Hwang}{Min
  et~al\mbox{.}}{2021}]%
        {min2021meta}
\bibfield{author}{\bibinfo{person}{Dongchan Min}, \bibinfo{person}{Dong~Bok
  Lee}, \bibinfo{person}{Eunho Yang}, {and} \bibinfo{person}{Sung~Ju Hwang}.}
  \bibinfo{year}{2021}\natexlab{}.
\newblock \showarticletitle{Meta-stylespeech: Multi-speaker adaptive
  text-to-speech generation}.
\newblock  (\bibinfo{year}{2021}), \bibinfo{pages}{7748--7759}.
\newblock


\bibitem[\protect\citeauthoryear{Oord, Li, Babuschkin, Simonyan, Vinyals,
  Kavukcuoglu, Driessche, Lockhart, Cobo, Stimberg, et~al\mbox{.}}{Oord
  et~al\mbox{.}}{2018}]%
        {oord2018parallel}
\bibfield{author}{\bibinfo{person}{Aaron Oord}, \bibinfo{person}{Yazhe Li},
  \bibinfo{person}{Igor Babuschkin}, \bibinfo{person}{Karen Simonyan},
  \bibinfo{person}{Oriol Vinyals}, \bibinfo{person}{Koray Kavukcuoglu},
  \bibinfo{person}{George Driessche}, \bibinfo{person}{Edward Lockhart},
  \bibinfo{person}{Luis Cobo}, \bibinfo{person}{Florian Stimberg},
  {et~al\mbox{.}}} \bibinfo{year}{2018}\natexlab{}.
\newblock \showarticletitle{Parallel wavenet: Fast high-fidelity speech
  synthesis}.
\newblock  (\bibinfo{year}{2018}), \bibinfo{pages}{3918--3926}.
\newblock


\bibitem[\protect\citeauthoryear{Oord, Dieleman, Zen, Simonyan, Vinyals,
  Graves, Kalchbrenner, Senior, and Kavukcuoglu}{Oord et~al\mbox{.}}{2016}]%
        {oord2016wavenet}
\bibfield{author}{\bibinfo{person}{Aaron van~den Oord}, \bibinfo{person}{Sander
  Dieleman}, \bibinfo{person}{Heiga Zen}, \bibinfo{person}{Karen Simonyan},
  \bibinfo{person}{Oriol Vinyals}, \bibinfo{person}{Alex Graves},
  \bibinfo{person}{Nal Kalchbrenner}, \bibinfo{person}{Andrew Senior}, {and}
  \bibinfo{person}{Koray Kavukcuoglu}.} \bibinfo{year}{2016}\natexlab{}.
\newblock \showarticletitle{Wavenet: A generative model for raw audio}.
\newblock \bibinfo{journal}{\emph{arXiv preprint arXiv:1609.03499}}
  (\bibinfo{year}{2016}).
\newblock


\bibitem[\protect\citeauthoryear{philsyn}{philsyn}{2021}]%
        {github2021DiffWaveVocoder}
\bibfield{author}{\bibinfo{person}{philsyn}.} \bibinfo{year}{2021}\natexlab{}.
\newblock \showarticletitle{DiffWave-Vocoder}.
\newblock
  \bibinfo{journal}{\emph{\url{https://github.com/philsyn/DiffWave-Vocoder}}}
  (\bibinfo{year}{2021}).
\newblock


\bibitem[\protect\citeauthoryear{Popov, Vovk, Gogoryan, Sadekova, and
  Kudinov}{Popov et~al\mbox{.}}{2021}]%
        {popov2021gradtts}
\bibfield{author}{\bibinfo{person}{Vadim Popov}, \bibinfo{person}{Ivan Vovk},
  \bibinfo{person}{Vladimir Gogoryan}, \bibinfo{person}{Tasnima Sadekova},
  {and} \bibinfo{person}{Mikhail Kudinov}.} \bibinfo{year}{2021}\natexlab{}.
\newblock \showarticletitle{Grad-TTS: A Diffusion Probabilistic Model for
  Text-to-Speech}.
\newblock  (\bibinfo{year}{2021}).
\newblock


\bibitem[\protect\citeauthoryear{Prenger, Valle, and Catanzaro}{Prenger
  et~al\mbox{.}}{2019}]%
        {prenger2019waveglow}
\bibfield{author}{\bibinfo{person}{Ryan Prenger}, \bibinfo{person}{Rafael
  Valle}, {and} \bibinfo{person}{Bryan Catanzaro}.}
  \bibinfo{year}{2019}\natexlab{}.
\newblock \showarticletitle{Waveglow: A flow-based generative network for
  speech synthesis}.
\newblock  (\bibinfo{year}{2019}), \bibinfo{pages}{3617--3621}.
\newblock


\bibitem[\protect\citeauthoryear{Protasio~Ribeiro, Florencio, Zhang, and
  Seltze}{Protasio~Ribeiro et~al\mbox{.}}{[n.d.]}]%
        {protasio_ribeiro_crowdmos_2011}
\bibfield{author}{\bibinfo{person}{Flavio Protasio~Ribeiro},
  \bibinfo{person}{Dinei Florencio}, \bibinfo{person}{Cha Zhang}, {and}
  \bibinfo{person}{Seltze}.} \bibinfo{year}{[n.d.]}\natexlab{}.
\newblock \showarticletitle{{CROWDMOS}: An Approach for Crowdsourcing Mean
  Opinion Score Studies}.
\newblock  (\bibinfo{year}{[n.\,d.]}).
\newblock
\newblock
\shownote{Edition: {ICASSP}.}


\bibitem[\protect\citeauthoryear{Ren, Hu, Tan, Qin, Zhao, Zhao, and Liu}{Ren
  et~al\mbox{.}}{2020a}]%
        {ren2020fastspeech}
\bibfield{author}{\bibinfo{person}{Yi Ren}, \bibinfo{person}{Chenxu Hu},
  \bibinfo{person}{Xu Tan}, \bibinfo{person}{Tao Qin}, \bibinfo{person}{Sheng
  Zhao}, \bibinfo{person}{Zhou Zhao}, {and} \bibinfo{person}{Tie-Yan Liu}.}
  \bibinfo{year}{2020}\natexlab{a}.
\newblock \showarticletitle{Fastspeech 2: Fast and high-quality end-to-end text
  to speech}.
\newblock \bibinfo{journal}{\emph{arXiv preprint arXiv:2006.04558}}
  (\bibinfo{year}{2020}).
\newblock


\bibitem[\protect\citeauthoryear{Ren, Ruan, Tan, Qin, Zhao, Zhao, and Liu}{Ren
  et~al\mbox{.}}{2019}]%
        {Fastspeech}
\bibfield{author}{\bibinfo{person}{Yi Ren}, \bibinfo{person}{Yangjun Ruan},
  \bibinfo{person}{Xu Tan}, \bibinfo{person}{Tao Qin}, \bibinfo{person}{Sheng
  Zhao}, \bibinfo{person}{Zhou Zhao}, {and} \bibinfo{person}{Tie-Yan Liu}.}
  \bibinfo{year}{2019}\natexlab{}.
\newblock \showarticletitle{Fastspeech: Fast, robust and controllable text to
  speech}.
\newblock  (\bibinfo{year}{2019}), \bibinfo{pages}{3171--3180}.
\newblock


\bibitem[\protect\citeauthoryear{Ren, Tan, Qin, Luan, Zhao, and Liu}{Ren
  et~al\mbox{.}}{2020b}]%
        {DeepSinger}
\bibfield{author}{\bibinfo{person}{Yi Ren}, \bibinfo{person}{Xu Tan},
  \bibinfo{person}{Tao Qin}, \bibinfo{person}{Jian Luan}, \bibinfo{person}{Zhou
  Zhao}, {and} \bibinfo{person}{Tie-Yan Liu}.}
  \bibinfo{year}{2020}\natexlab{b}.
\newblock \showarticletitle{Deepsinger: Singing voice synthesis with data mined
  from the web}.
\newblock  (\bibinfo{year}{2020}), \bibinfo{pages}{1979--1989}.
\newblock


\bibitem[\protect\citeauthoryear{Rix, Beerends, Hollier, and Hekstra}{Rix
  et~al\mbox{.}}{2001}]%
        {rix2001perceptual}
\bibfield{author}{\bibinfo{person}{Antony~W Rix}, \bibinfo{person}{John~G
  Beerends}, \bibinfo{person}{Michael~P Hollier}, {and}
  \bibinfo{person}{Andries~P Hekstra}.} \bibinfo{year}{2001}\natexlab{}.
\newblock \showarticletitle{Perceptual evaluation of speech quality (PESQ)-a
  new method for speech quality assessment of telephone networks and codecs}.
\newblock  (\bibinfo{year}{2001}).
\newblock


\bibitem[\protect\citeauthoryear{Shen, Pang, Weiss, Schuster, Jaitly, Yang,
  Chen, Zhang, Wang, Skerrv-Ryan, et~al\mbox{.}}{Shen et~al\mbox{.}}{2018}]%
        {tacotron}
\bibfield{author}{\bibinfo{person}{Jonathan Shen}, \bibinfo{person}{Ruoming
  Pang}, \bibinfo{person}{Ron~J Weiss}, \bibinfo{person}{Mike Schuster},
  \bibinfo{person}{Navdeep Jaitly}, \bibinfo{person}{Zongheng Yang},
  \bibinfo{person}{Zhifeng Chen}, \bibinfo{person}{Yu Zhang},
  \bibinfo{person}{Yuxuan Wang}, \bibinfo{person}{Rj Skerrv-Ryan},
  {et~al\mbox{.}}} \bibinfo{year}{2018}\natexlab{}.
\newblock \showarticletitle{Natural tts synthesis by conditioning wavenet on
  mel spectrogram predictions}.
\newblock  (\bibinfo{year}{2018}), \bibinfo{pages}{4779--4783}.
\newblock


\bibitem[\protect\citeauthoryear{Taal, Hendriks, Heusdens, and Jensen}{Taal
  et~al\mbox{.}}{2010}]%
        {taal2010short}
\bibfield{author}{\bibinfo{person}{Cees~H Taal}, \bibinfo{person}{Richard~C
  Hendriks}, \bibinfo{person}{Richard Heusdens}, {and} \bibinfo{person}{Jesper
  Jensen}.} \bibinfo{year}{2010}\natexlab{}.
\newblock \showarticletitle{A short-time objective intelligibility measure for
  time-frequency weighted noisy speech}.
\newblock  (\bibinfo{year}{2010}).
\newblock


\bibitem[\protect\citeauthoryear{Wang, Takaki, and Yamagishi}{Wang
  et~al\mbox{.}}{2019}]%
        {wang2019neural}
\bibfield{author}{\bibinfo{person}{Xin Wang}, \bibinfo{person}{Shinji Takaki},
  {and} \bibinfo{person}{Junichi Yamagishi}.} \bibinfo{year}{2019}\natexlab{}.
\newblock \showarticletitle{Neural source-filter waveform models for
  statistical parametric speech synthesis}.
\newblock \bibinfo{journal}{\emph{IEEE/ACM Transactions on Audio, Speech, and
  Language Processing}}  \bibinfo{volume}{28} (\bibinfo{year}{2019}),
  \bibinfo{pages}{402--415}.
\newblock


\bibitem[\protect\citeauthoryear{Yamamoto, Song, and Kim}{Yamamoto
  et~al\mbox{.}}{2020}]%
        {yamamoto2020parallel}
\bibfield{author}{\bibinfo{person}{Ryuichi Yamamoto}, \bibinfo{person}{Eunwoo
  Song}, {and} \bibinfo{person}{Jae-Min Kim}.} \bibinfo{year}{2020}\natexlab{}.
\newblock \showarticletitle{Parallel WaveGAN: A fast waveform generation model
  based on generative adversarial networks with multi-resolution spectrogram}.
\newblock  (\bibinfo{year}{2020}), \bibinfo{pages}{6199--6203}.
\newblock


\bibitem[\protect\citeauthoryear{Yang, Yang, Liu, Fang, Chen, and Xie}{Yang
  et~al\mbox{.}}{2020}]%
        {yang2020multiband}
\bibfield{author}{\bibinfo{person}{Geng Yang}, \bibinfo{person}{Shan Yang},
  \bibinfo{person}{Kai Liu}, \bibinfo{person}{Peng Fang}, \bibinfo{person}{Wei
  Chen}, {and} \bibinfo{person}{Lei Xie}.} \bibinfo{year}{2020}\natexlab{}.
\newblock \showarticletitle{Multi-band MelGAN: Faster Waveform Generation for
  High-Quality Text-to-Speech}.
\newblock  (\bibinfo{year}{2020}).
\newblock
\showeprint[arxiv]{2005.05106}~[cs.SD]


\bibitem[\protect\citeauthoryear{Zhang, Tan, Yu, Zhao, Kuang, Liu, Zhou, Yang,
  and Wu}{Zhang et~al\mbox{.}}{2020a}]%
        {DBLP:conf/mm/ZhangTYZKLZYW20}
\bibfield{author}{\bibinfo{person}{Shengyu Zhang}, \bibinfo{person}{Ziqi Tan},
  \bibinfo{person}{Jin Yu}, \bibinfo{person}{Zhou Zhao}, \bibinfo{person}{Kun
  Kuang}, \bibinfo{person}{Jie Liu}, \bibinfo{person}{Jingren Zhou},
  \bibinfo{person}{Hongxia Yang}, {and} \bibinfo{person}{Fei Wu}.}
  \bibinfo{year}{2020}\natexlab{a}.
\newblock \showarticletitle{Poet: Product-oriented Video Captioner for
  E-commerce}. In \bibinfo{booktitle}{\emph{{MM} '20: The 28th {ACM}
  International Conference on Multimedia}}. \bibinfo{publisher}{{ACM}},
  \bibinfo{pages}{1292--1301}.
\newblock


\bibitem[\protect\citeauthoryear{Zhang, Tan, Zhao, Yu, Kuang, Jiang, Zhou,
  Yang, and Wu}{Zhang et~al\mbox{.}}{2020b}]%
        {DBLP:conf/kdd/ZhangTZYKJZYW20}
\bibfield{author}{\bibinfo{person}{Shengyu Zhang}, \bibinfo{person}{Ziqi Tan},
  \bibinfo{person}{Zhou Zhao}, \bibinfo{person}{Jin Yu}, \bibinfo{person}{Kun
  Kuang}, \bibinfo{person}{Tan Jiang}, \bibinfo{person}{Jingren Zhou},
  \bibinfo{person}{Hongxia Yang}, {and} \bibinfo{person}{Fei Wu}.}
  \bibinfo{year}{2020}\natexlab{b}.
\newblock \showarticletitle{Comprehensive Information Integration Modeling
  Framework for Video Titling}. In \bibinfo{booktitle}{\emph{{KDD} '20: The
  26th {ACM} {SIGKDD} Conference on Knowledge Discovery and Data Mining}}.
  \bibinfo{publisher}{{ACM}}, \bibinfo{pages}{2744--2754}.
\newblock


\bibitem[\protect\citeauthoryear{Zhang, Yang, Yao, Lu, Feng, Zhao, Chua, and
  Wu}{Zhang et~al\mbox{.}}{2022}]%
        {DBLP:conf/www/ZhangYYLFZC022}
\bibfield{author}{\bibinfo{person}{Shengyu Zhang}, \bibinfo{person}{Lingxiao
  Yang}, \bibinfo{person}{Dong Yao}, \bibinfo{person}{Yujie Lu},
  \bibinfo{person}{Fuli Feng}, \bibinfo{person}{Zhou Zhao},
  \bibinfo{person}{Tat{-}Seng Chua}, {and} \bibinfo{person}{Fei Wu}.}
  \bibinfo{year}{2022}\natexlab{}.
\newblock \showarticletitle{Re4: Learning to Re-contrast, Re-attend,
  Re-construct for Multi-interest Recommendation}. In
  \bibinfo{booktitle}{\emph{{WWW} '22: The {ACM} Web Conference 2022}}.
  \bibinfo{publisher}{{ACM}}, \bibinfo{pages}{2216--2226}.
\newblock


\end{thebibliography}

\appendix
\clearpage
\section{Architecture} \label{architecture}


In this section, we list the architecture hyperparameters of SingGAN in Table \ref{tab:hyperparameters_gen} and \ref{tab:hyperparameters_dis}, and the architecture hyperparameters of SingGAN-SVS in Table \ref{tab:hyperparameters_SVS}.

\begin{table}[h]
\centering
\begin{tabular}{l|c}
\toprule
\textbf{Hyperparameter}            &  \textbf{SingGAN}  \\
\midrule
Number of Harmonic                &     8     \\ 
Generator Kernel Size             &     5     \\ 
Number of AFL blocks              &     3    \\           
Layers Per AFL blocks             &     10    \\          
Generator Residual Channels          &     64    \\
Generator Upsample Scales             &    [8, 4, 4]  \\   
\midrule
Total Number of Parameters      &       1.59M	         \\ 
\bottomrule
\end{tabular}
\caption{Architecture hyperparameters of Generator.}
\vspace{-5mm}
\label{tab:hyperparameters_gen}
\end{table}

\begin{table}[h]
\centering
\begin{tabular}{l|c}
\toprule
\textbf{Hyperparameter}            &  \textbf{SingGAN}  \\
\midrule
Kernel Size                                      &     3     \\       
Residual Channels                              &     64    \\              
Full-band Discriminator Number of Layers        &     10    \\
Number of Sub-band Discriminators               &     4    \\
Sub-band Discriminator Kernel Size             &     [5, 5, 7, 7]     \\                     
Sub-band Discriminator Number of Layers        &     [8, 8, 6, 6]    \\
LeakyReLU Activation Slope                     &     0.2    \\
\midrule
Total Number of Parameters        &       0.5M	         \\ 
\bottomrule
\end{tabular}
\caption{Architecture hyperparameters of Discriminator.}
\vspace{-5mm}
\label{tab:hyperparameters_dis}
\end{table}

\begin{table}[h]
\centering
\begin{tabular}{l|c}
\toprule
\textbf{Hyperparameter}   & \textbf{SingGAN-SVS} \\ 
\midrule
Phoneme Embedding           &256   \\
Text Encoder Layers              &4   \\
Text Encoder Hidden              &256     \\                      
Text Encoder Conv1d Kernel       &9   \\    
Text Encoder Conv1D Filter Size  &1024\\                 
Text Encoder Attention Heads     &2   \\    
Text Encoder Dropout             &0.05\\                       
\midrule
Variance Predictor Conv1D Kernel        & 3\\    
Variance Predictor Conv1D Filter Size   & 256  \\    
Variance Predictor Dropout              & 0.5  \\ 
\midrule
Decoder Layers                      &   4 \\    
Decoder Hidden                      &   256  \\    
Decoder Conv1D Kernel               &    9   \\    
Decoder Conv1D Filter Size          &  1024  \\   
Decoder Attention Headers           &  2    \\       
Decoder Dropout                     &  0.1  \\  
\midrule
Total Number of Parameters   & 28M  \\
\bottomrule
\end{tabular}
\caption{Architecture hyperparameters of the Lyric-to-spectrogram generation model in SingGAN-SVS.}
\vspace{-5mm}
\label{tab:hyperparameters_SVS}
\end{table}

\section{Loss Details} \label{loss}
Here we introduce the hyperparameter details in the auxiliary spectrogram losses.

\begin{table}[H]
  \centering
  \begin{tabular}{cccc}
  \toprule
  Loss type             & FFT Size & Hop Size & Win Length \\
  \midrule
 \multirow{2}{*}{Mel-spectrogram} & 2048   & 270   & 1080      \\
                                & 4096  & 540    & 2160      \\
 \midrule
 \multirow{3}{*}{STFT}   & 512       & 50         & 240      \\
                         & 1024     & 120         & 600        \\
                        & 2048     & 240         & 1200        \\
 \bottomrule
 \end{tabular}
 \vspace{2mm}
\caption{The details of the multi-resolution STFT loss and multi-resolution mel-spectrogram loss. A Hanning window is applied before the FFT process.}
\label{table:spectrogram_loss}
\end{table}

\section{Evaluation Matrix} \label{appendix:evaluation}

\subsection{Subjective Evaluation}
All our Mean Opinion Score (MOS) tests are crowdsourced and conducted by native speakers. We refer to the rubric for MOS scores in~\cite{protasio_ribeiro_crowdmos_2011}, and the scoring criteria has been included in Table~\ref{matrix:naturalness} for completeness. The samples are presented and rated one at a time by the testers. 

\begin{table}[H]
  \vspace{-2mm}
  \begin{tabular}{ccc}
  \toprule
  Rating & Naturalness & Definition                           \\
  \midrule
  1      & Bad        &  Very annoying and objectionable dist. \\
  2      & Poor       &  Annoying but not objectionable dist. \\
  3      & Fair       &  Perceptible and slightly annoying dist\\
  4      & Good       & Just perceptible but not annoying dist. \\
  5      & Excellent  & Imperceptible distortions\\
  \bottomrule
  \end{tabular}
  \caption{Ratings that have been used in evaluation of speech naturalness of synthetic and ground truth samples.}
  \vspace{-4mm}
    \label{matrix:naturalness}
\end{table}

\subsection{Objective Evaluation} 
Perceptual evaluation of speech quality (PESQ)~\cite{rix2001perceptual} and the short-time objective intelligibility (STOI)~\cite{taal2010short} assesses the denoising quality for speech enhancement.

Mel-cepstral distortion (MCD)~\cite{kubichek1993mel} measures the spectral distance between the synthesized and reference mel-spectrum features.

\section{Speech Synthesis}
  To further verify the generality of SingGAN, we conduct experiments on the multi-speaker English dataset VCTK and the single-speaker Chinese dataset CSMSC.
  Our configuration follows prior singing voice synthesis systems. We perform MOS tests and present the results in Table~\ref{table:mos5}.
  The results show that SingGAN achieves good results in speech synthesis as well. 
  Although MOS differences in speech are not so much as that in the singing voice, it is obvious that SingGAN outperforms prior neural vocoders in audio synthesis.
  Consequently, our proposed SingGAN has the potential to be widely applied in TTS and SVS.

  \begin{table}[h]
    \centering
    \begin{tabular}{ccc}
      \toprule
    \multirow{2}{*}{\bfseries Model} & \multicolumn{2}{c}{\bfseries Speech MOS} \\
                           & \bfseries English  & \bfseries Mandarin     \\
    \midrule
    WaveRNN                &  4.02$\pm$0.14           &  4.24$\pm$0.10            \\
    FB-MelGAN              &  3.62$\pm$0.09           &  3.86$\pm$0.10            \\
    Parallel WaveGAN       &  4.12$\pm$0.08           &  4.20$\pm$0.09            \\
    SingGAN            &  4.26$\pm$0.10           &  4.32$\pm$0.09            \\
    \bottomrule
    \end{tabular}
    \caption{The MOS results with 95\% confidence intervals on the English VCTK and Chinese CSMSC dataset.}
    \label{table:mos5}
    \end{table}

\clearpage
\end{document}